\documentclass[a4paper,journal,onecolumn,12pt]{IEEEtran}
\usepackage{amsmath,amsfonts}
\usepackage{algorithmic}
\usepackage{algorithm}
\usepackage{array}
\usepackage[caption=false,font=normalsize,labelfont=sf,textfont=sf]{subfig}
\usepackage{textcomp}
\usepackage{stfloats}
\usepackage{url}
\usepackage{verbatim}
\usepackage{graphicx}
\usepackage{cite}
\usepackage[a4paper,left=3cm,right=3cm,top=3cm,bottom=3cm]{geometry}

\usepackage{bm}

\usepackage{lineno}

\usepackage{listings}%

\usepackage{todonotes}

\usepackage{pdfpages}
\usepackage{setspace}

\newtheorem{lemma}{Lemma}
\newtheorem{theorem}[lemma]{Theorem}%

\begin{document}


\title{Non-Random Data Encodes Its Geometric and Topological Dimensions}

\author{Hector Zenil, Felipe S. Abrah\~ao, Luan Ozelim

\thanks{(HZ) School of Biomedical Engineering and Imaging Sciences, King’s College London, U.K.}

\thanks{(HZ) The Alan Turing Institute, British Library, U.K.}

\thanks{(HZ) King’s Institute for Artificial Intelligence, King’s College London, U.K.}

\thanks{(HZ, LO) The Arrival Institute, U.K.}

\thanks{(HZ, FSA, LO) Oxford Immune Algorithmics, Oxford University Innovation, U.K.}

\thanks{(HZ, FSA, LO) Algorithmic Dynamics Lab, Center for Molecular Medicine, Karolinska Institutet, Sweden.}

\thanks{(FSA) Centre for Logic, Epistemology and the History of Science, University of Campinas (UNICAMP), Brazil.}

\thanks{(FSA) DEXL, National Laboratory for Scientific Computing (LNCC), Brazil.}

\thanks{Corresponding author: hector.zenil@kcl.ac.uk}

}



\maketitle

\begin{abstract}
Based on the principles of information theory, measure theory, and theoretical computer science, we introduce a signal deconvolution method with a wide range of applications to coding theory, particularly in zero-knowledge one-way communication channels, such as in deciphering messages (i.e., objects embedded into multidimensional spaces) from unknown generating sources about which no prior knowledge is available and to which no return message can be sent. Our multidimensional space reconstruction method from an arbitrary received signal is proven to be agnostic vis-\`a-vis the encoding-decoding scheme, computation model, programming language, formal theory, the computable (or semi-computable) method of approximation to algorithmic complexity, and any arbitrarily chosen (computable) probability measure. The method derives from the principles of an approach to Artificial General Intelligence (AGI) capable of building a general-purpose model of models independent of any arbitrarily assumed prior probability distribution. We argue that this optimal and universal method of decoding non-random data has applications to signal processing, causal deconvolution, topological and geometric properties encoding, cryptography, and bio- and technosignature detection.
\end{abstract}


\onehalfspacing
\section{Introduction}

All of our knowledge about biological systems is currently based on what occurs here on Earth~\cite{zenillife,bedauOpenProblemsArtificial2000, clelandDefiningLife2002, dupreMetaphysicsBiology2021, kempesMultiplePathsMultiple2021, mariscalHiddenConceptsHistory2019, ruiz-mirazoUniversalDefinitionLife2004, walkerAlgorithmicOriginsLife2013, witzanyWhatLife2020}. In order to explore new scenarios in which life and intelligence may appear, in addition to discovering signs or traces of extraterrestrial biological systems, some researchers consider technosignature detection a more promising method of detecting life than other biosignatures because of their longevity~\cite{zengauch,haqq-misraSearchingTechnosignaturesExoplanetary2022a, wrightCaseTechnosignaturesWhy2022}.
However, some researchers worry that our current theoretical toolkit for data analysis may not be  sophisticated enough~\cite{bottaStrategiesLifeDetection2008, enyaComparativeStudyMethods2022, neveuLadderLifeDetection2018}. 
%
%
Most of the discussion has been centred around the technical justification of hardware choices and the technicalities of detection (e.g. the frequency) and not the nature (i.e., original structural characteristics or underlying mathematical properties) of the signal itself. 
However, 
beyond identifying very basic statistical patterns, not much progress has been made on the qualitative semantic aspects of signals and communication, 
let alone on building a universal framework within which the question can be explored and analysed. 

There are myriad data types and formats that encode different types of data on multiple levels, from ASCII characters to video files to health record data. While it is possible to re-encode all data into a single format, the original message may be lost (imagine trying to extract voice-recorded messages of a sound file only by looking at the sound wave plot over time).
Because of this diversity even within our own technology, the process of detecting, decoding, and interpreting terrestrial or extraterrestrial signals \textit{must} involve considering a wide range of possible encoding and decoding schemes.


Without proper conditions that assure the decoding algorithms can only pick up meaningful or sound features---given the prior assumption that a particular dataset actually contains a message---we know the presence of spurious correlations constitutes one of the dangers of naively applying any sort of statistics-based Big Data analysis to extract or find statistically significant patterns in sufficiently large datasets~\cite{Calude2017,Smith2020}. 
Ideally, given a random signal such as white noise, no such signal interpretation should yield any meaningful message or interpretation.
A random signal should not only look random, but also be interpreted as such, no matter how it is decoded, or how it was encoded.
Thus, along with the data compression distortions known to occur from the application of statistics-based or entropy-based methods~\cite{zkpaper,zenilreview2020}, the possibility of spurious findings in random messages is of major concern in decoding signal streams so as to reconstruct the original message.
Unless the receiver agent has
prior knowledge about the encoding-decoding scheme beforehand, such an incompressible message would be indistinguishable from an ideal fair-coin-toss sequence of events.
Moreover, notice that it is the receiver's
prior knowledge about how to decode the signal according to the emitter's previously chosen encoding scheme what 
guarantees the validity of
the classical noisy-channel coding theorem~\cite{Cover2005}. 
It demonstrates that the emitter agent can introduce redundancies in the original message---thus making the signal data more compressible or non-random---to be correctly interpreted by the receiver so as to avoid decoding noise as meaningful information.

\emph{One-way communication} channels are those for which the receiver cannot (in principle or in practice) send any signal back to the emitter in order to help or facilitate the decoding process of the first message sent by the emitter. 
The mathematical phenomena discussed in the above paragraph strongly suggest that for effective communication, especially for one-way communication between any emitter agent and receiver agent, a message should \emph{not} be displayed in its most compressed form, unless the receiver has knowledge of the compression schema and can uncompress the message back into its original state (disregarding efficiency).  
Instead, a message should be sent and received in an uncompressed form.

In this case in which the message is encoded in an uncompressed form, or equivalently in which the message is compressible (therefore non-random), one would have a large standard deviation in the range of information/complexity values for the different ways of decomposing, parsing or perturbing the signal. 
(See Section~\ref{sectionTheory} and Appendix~\ref{sectionAIDistortions}).
Thus, the schema that shortens any redundancies must be read properly by an interpreter to accurately translate the shorthand redundancies back into their longer form.
Indeed, it is possible for a signal stream to arrive at a naive receiver in a format that contains sufficient information about how it ought to be decompressed. 
However, even in this case, the receiver must have prior knowledge  about how to decode that signal into its original encoding so as to reveal the redundancies present in the original message that gave it meaning.
This prior knowledge means, for example, that one previously knows the chosen encoding-decoding scheme, the programming language, computation model, or the formal theory that explains the relevant physical phenomena which the emitter agent is trying to convey.

In the face of such theoretical challenges, the problem that we tackle in this article is how to encode/decode the message in a compressible signal so as to enable the correct decoding without prior knowledge of the chosen encoding-decoding scheme.
We introduce a method based upon the principles of (algorithmic) information theory that is aimed at sweeping over various possible encoding and decoding schemes to test our current limits on signal interpretation as we attempt to reconstruct the original \emph{partition} (or in general, the \emph{multidimensional space} or \emph{structure}) in which the original message was given its ``meaning'' by the emitter agent.
This \emph{semantic} characteristic refers to the (algorithmic) information about the context or real-world correspondents of the emitter that the signal sent by this emitter is trying to convey to the receiver.
The real-world correspondents that we particularly investigate are the objects grounded in their respective \emph{contexts}, that is, embedded into their respective multidimensional spaces (or structures) as the emitter originally intended.

We demonstrate how messages may be reconstructed by deriving the number of dimensions and the scale of each length of an object from examples ranging from text to images embedded in multiple dimensions, showing a connection between irreducible information content, syntax, geometry, semantics, and topology.
This article also shows results that are agnostic vis-a-vis prior knowledge of encoding-decoding schemes, and also demonstrate sufficient conditions in this zero-knowledge scenario for enabling the reconstruction of the original \emph{message} (i.e., the original object embedded into the original multidimensional space to be conveyed to the receiver) in one-way communication channels.

\emph{Zero-knowledge communication} occurs when the receiver agent is able to correctly interpret the received signal as the originally intended message sent by the emitter agent, given that the receiver has no knowledge about the encoding-decoding scheme chosen by the emitter.
Notice that this condition of no prior information about the emitter agent only applies to encoding-decoding schemes---in addition to other language, computational, and mathematical modeling characteristics, such as the programming language, computational model, and probability measure---, and (consequentially) also to the original multidimensional space and object that are unknown to the receiver before any communication takes place.
However, as discussed in the Supplementary Material provided along with the present article, this zero-knowledge condition does not mean other assumptions regarding the unknown emitter agent are not being considered by the receiver agent.
Also, the reader should not confuse zero-knowledge communication (ZKC) with zero-knowledge proof (ZKP), which is commonly studied in cryptography~\cite{Buchanan2022Cryptographybook,Vadhan2023surveyonZKproofs,Allender2023AITstatisticalZK}.
Actually, 
one may consider ZKP and ZKC as akin mathematical problems but as diametrically opposing counterparts regarding the acquisition of knowledge (see also Appendix~\ref{sectionZKC}).

The method is based on the principles of Algorithmic Information Dynamics (AID)~\cite{algodyn,nmi,Abrahao2021b}, and it consists of a perturbation analysis of a received signal. 
AID is based upon the principles of information theory and the mechanisms of algorithmic probability and the universal distribution, a formal approach to a type of Artificial General/Super Intelligence that requires the massive production of a universal distribution, the mother of all models~\cite{miracle}, to build a very large (semi-computable) model of computable approximate models.  
The underlying idea is that a computable model is a causal explanation of a piece of data.  
As a result of each perturbation applied to the received signal, a new computable approximate model is built, and then compared against the observation~\cite{nmi,maininfo}.

As explained in Section~\ref{sectionMethods} and formalised in the Sup. Mat., we estimate the algorithmic probability changes for a message under distinct partitions that result from perturbations.
This process then builds a large landscape of possible candidate models (along with their respective approximate complexity values) from which one can infer the best model.
We show that the partition (i.e., multidimensional space) for which the message displays the lowest complexity indicates the original partition that the received signal stream encodes.
For example in the case of bidimensional spaces, our method consists of finding the bidimensional space $ \mathcal{ S }_2 $ for which the algorithmic complexity of the object embedded into $ \mathcal{ S }_2 $ (i.e., the algorithmic complexity of the message) is minimised in comparison to the landscape of other complexity values calculated from a sufficiently large number of perturbations on the received signal.


In this manner, our present approach also contributes a universal method that is not reducible to a straightforward application of algorithmic information theory (AIT).
As the number of perturbed messages strongly dominates the number of possible best message candidates, our theoretical and empirical results show that perturbation analysis enables the downward spikes displayed in the landscapes of complexities to be more likely to indicate the best message candidates, even though the exact values of the algorithmic complexities are semi-computable (i.e., these values are uncomputable and one can only asymptotically approximate them from above).
Although there is this intrinsically observer-dependent ``deficiency'' in the perturbation analysis that forbids an arbitrarily precise approximation to algorithmic complexity, we demonstrate that this subjective aspect can be eventually cancelled out in the perturbation analysis phase.
In fact, it is this theoretical limitation given by semi-computability that, counter-intuitively, enables the perturbation analysis to overcome this and the other subjective aspects (such as an arbitrarily chosen programming language, probability measure, or the computational appliances available) involved in the signal processing and interpretation process---that is, in finding an objective piece of information.



\section{Empirical Results}\label{sectionResults}

\subsection{On the perturbation analysis of compressed data}\label{sectionPerturbationAnalysis}

Our (structural) perturbation method consists in sweeping over a large variety of possible dimensional configurations and measuring the resulting information content and algorithmic complexity of the candidate messages by taking the lowest algorithmic complexity configurations under the assumption that the original message is algorithmically not random (i.e., that the original message is compressible).  
As shown in~\cite{Zenil2019b} for graphs, this is because if the original message is random, then the likelihood that a rearrangement (i.e., a new partition) will lead to a low complexity configuration is very low. On the other hand, if the original message is sufficiently compressible (see Sections~\ref{sectionTheory} and~\ref{sectionTowardgeneralZKC}), and therefore not random, most partitions will lead to configurations of greater algorithmic randomness.

As shown in~\cite{Zenil2024ETpaper1} which applies the method discussed in this article to reconstruct technosignatures, drops (i.e., downward spikes) in the complexity landscapes indicate candidate dimensions for their respective original encoding dimensions, and thus the decoding dimension that would process the signal such that it produces the lowest-complexity message.
That is, the original encoding of the message (the one with an interpretable meaning) is the one with the lowest complexity.
Our results indicate that sequences that carry meaningful content are those whose complexity quantitatively diverges from the vast majority of the other perturbed messages according to complexity measures.
The number of deviations (i.e., perturbations) from the encoded messages that cause the randomness to increase dominates those that do not.
In this scenario where a sufficiently large amount of perturbations is taken into account, our results show that an incoming signal with low randomness is more likely to encode a message than one that has more randomness.


In particular, the Block Decomposition Method (BDM)~\cite{bdmpaper} (combining Shannon entropy for long ranges and algorithmic probability for short ranges) was shown in~\cite{Zenil2024ETpaper1} to be the complexity measure that was most sensitive and accurate in identifying the original message (image) across the complexity landscapes.


Additionally, 
BDM significantly outperforms Compression and Shannon Entropy in the face of additive noise.
As shown in~\cite{Zenil2024ETpaper1},
algorithmic complexity (as estimated via BDM) remains invariant to low-algorithmic-complexity transformations that preserve the algorithmic content of the data (such as swapping all 0s and 1s or swapping all black and white pixels in an image), as well as some non-linear but algorithmic-information-content-preserving transformations~\cite{kolmo2d,bdmpaper}.


While the above discussed cases study the bidimensional spaces, Fig.~\ref{bone} below shows that our method can be scaled up to tridimensional spaces.
As we will present in Section~\ref{sectionTheory} (and it is demonstrated in the Sup. Mat.), we have found sufficient conditions for which the method remains sound for any multidimensional space.
Future research and challenges to tackle this generalization are fully explained in 
Appendices~
\ref{sectionAIDistortions}--\ref{sectionAlgorithmiclikelihood}.

\begin{figure}[H]
	\flushleft\footnotesize{\hspace{3cm}128 $\times$ 128
		$\times$ 128 \hspace{3.7cm} 64 $\times$ 128 $\times$  192}
	\centerline{\includegraphics[scale=0.28]{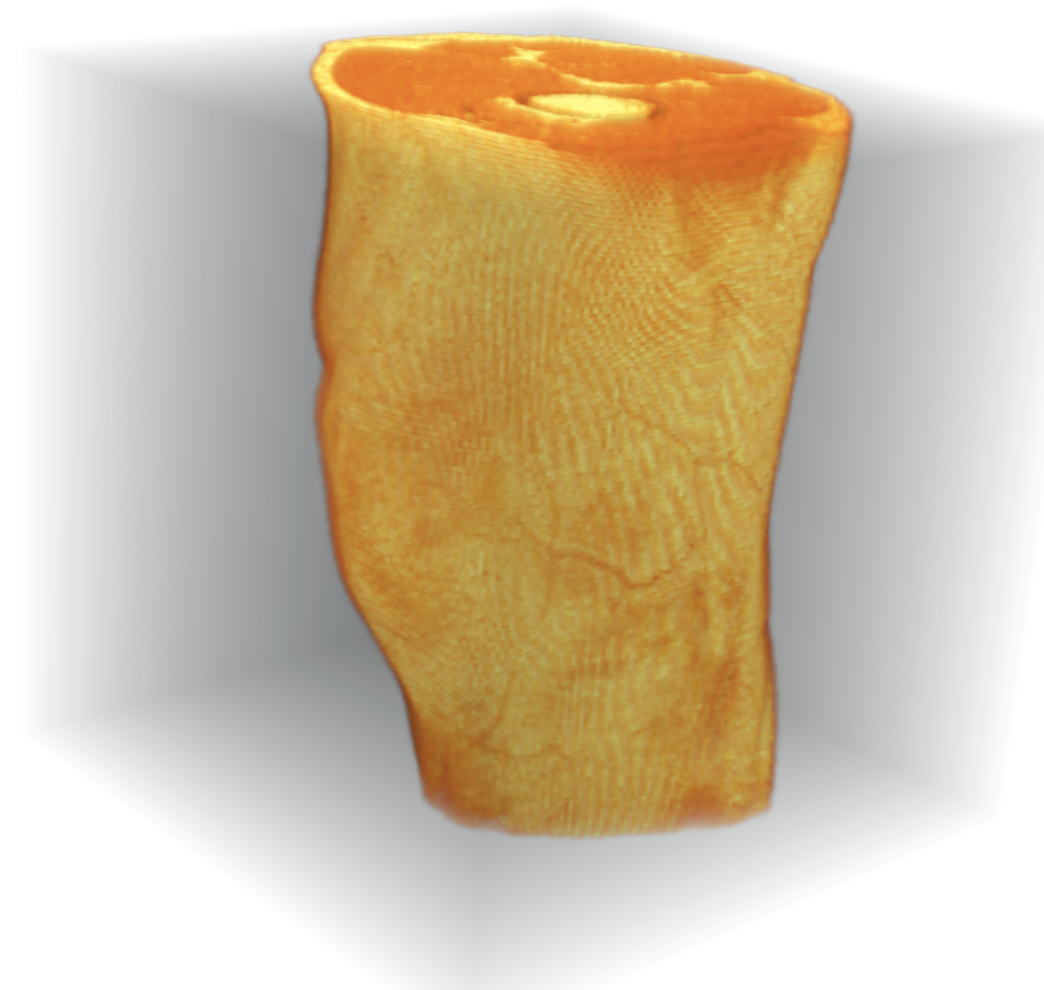}\hspace{1.5cm}\includegraphics[scale=0.27]{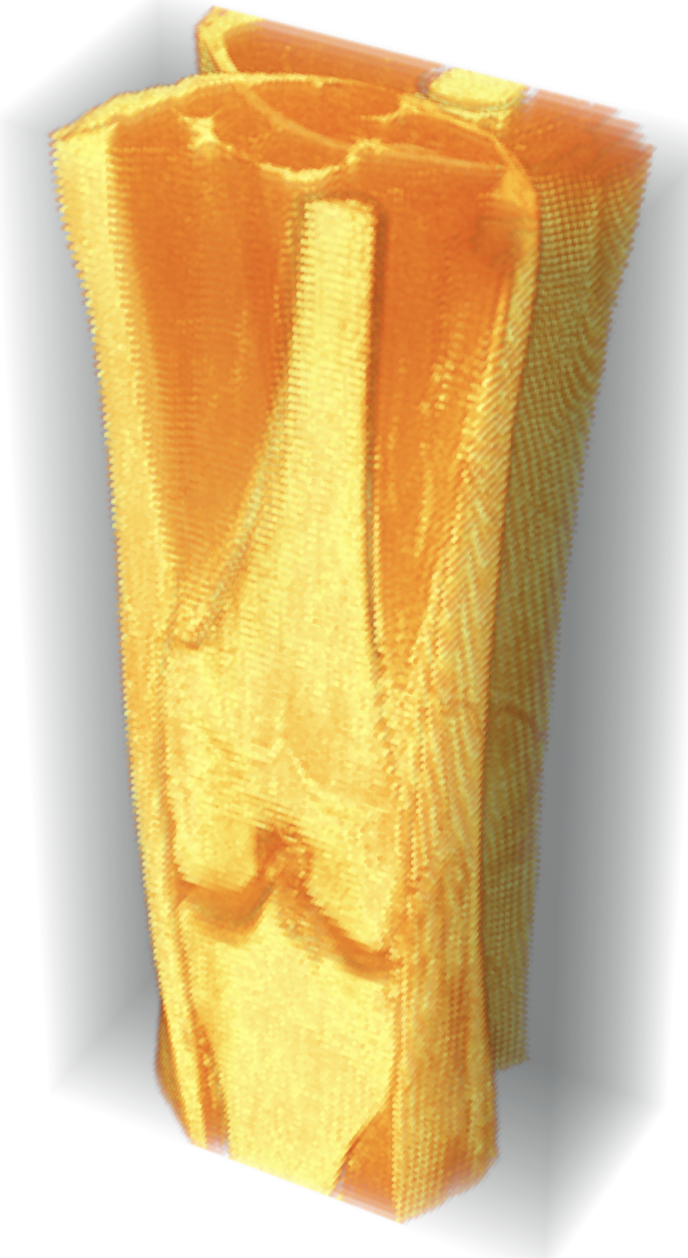}}
	\centerline{\includegraphics[scale=0.35]{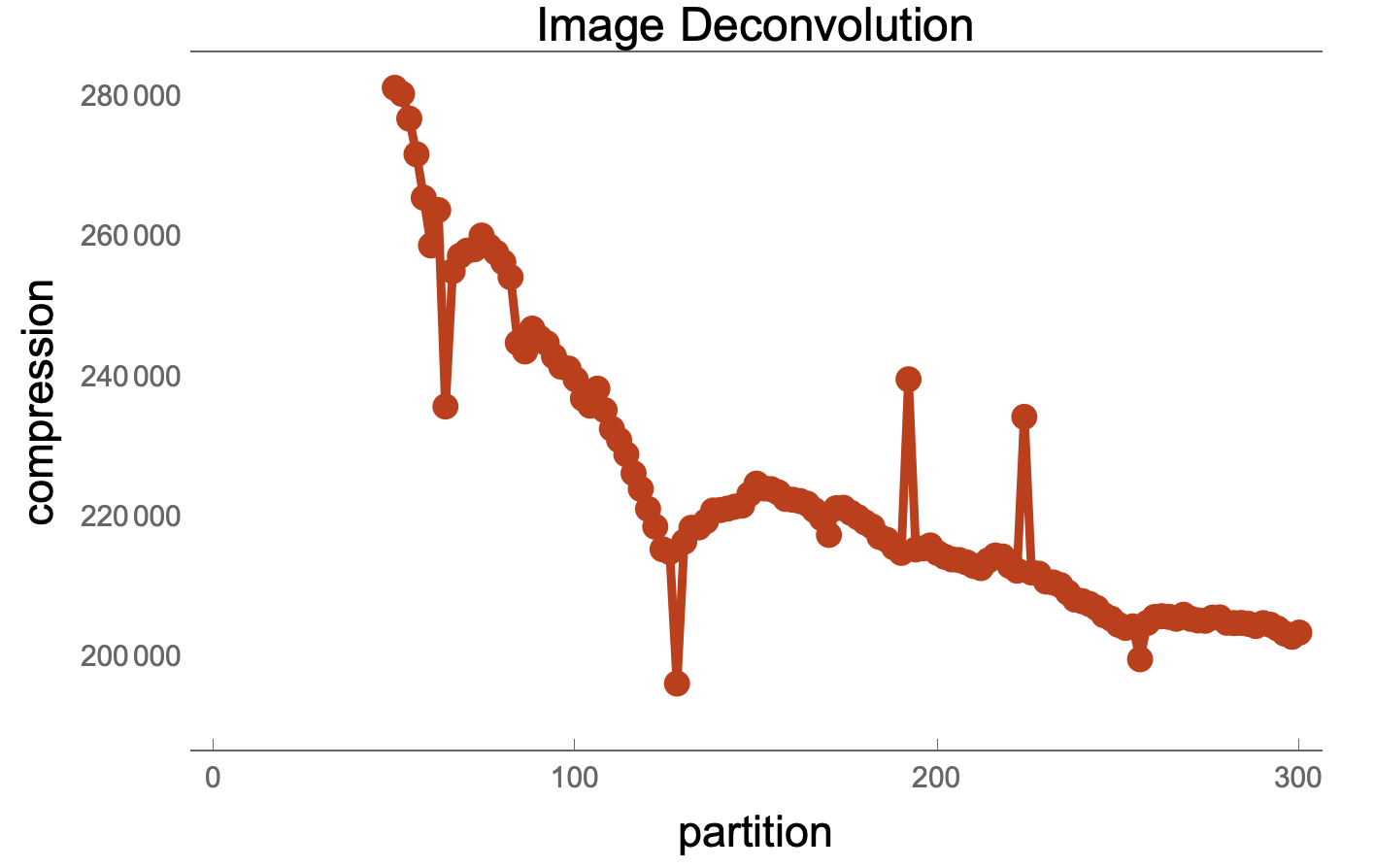}}
	\caption{\label{bone}\textit{Top left:} A reconstruction exercise of a 3D image of a Magnetic Resonance image of a knee embedded in a cube. \textit{Top right:} Reconstruction from the \textit{bottom:} perturbation analysis on various partitions. Spikes occur at the original dimension's multiples: 64, 128, and 192. When the linear signal stream is partitioned at the first candidate, the next dimensions are indicated by downward spikes, or upward spikes even on the first pass. A mirror image (top right) is indicated and reconstructed as the most likely candidate, and the correct knee configuration appears at the second spike (top left).}
\end{figure}

The next subsection presents novel empirical evidence of the suitability of not only structural perturbations, but also local bit-wise operations, as a robust strategy to find the best encode/decode scenario for a given message.

\subsection{Information content and bitwise perturbations}\label{sectionAIDonDarwins}

In introducing the bit-wise perturbation method put forward in this work and also illustrating the changes to compression, entropy and algorithmic complexity when considering different encoding methods, we analyse three types of strings:

\begin{itemize}
    \item A random binary string;
    \item A string of random characters;
    \item Real sentences from Darwin's ``Origin of Species''
\end{itemize}

The first step in the analysis is to bring all these strings into a common format. Thus, binary encodings were chosen such that both the strings of random characters and the sentences from Darwin's book were binarised in a few different ways according to arbitrary UTF8-to-binary functions. 

The proposed method is based on the premise that perturbations (with respect to both individual bits and spatial rearrangements of higher dimensional data) give us hints about the correct information content (including geometrical features) of a message. In order to assess the information content change, four metrics are measured for each perturbed string:

\begin{itemize}
    \item LZW: The length of the dictionary produced after applying a simple LZW lossless compression scheme to the input string;
    \item Entropy: Shannon entropy value of the string (which could be Block Entropy for higher dimensional data);
    \item BDM: The Block Decomposition Method (BDM)~\cite{bdmpaper} is a strategy used to approximate the algorithmic or algorithmic complexity of objects like strings and graphs. Unlike conventional methods that focus on compressibility, such as lossless compression techniques, BDM divides a complex object into smaller, easier-to-handle components. These individual blocks are then evaluated based on their complexity using a reference database of small Turing machine outputs. This database provides a baseline for assessing the algorithmic content of each block. BDM builds upon the Coding Theorem Method (CTM), which relates algorithmic probability to algorithmic complexity by determining the chance that a random program could produce a given string on a universal Turing machine. By applying this logic to each block and summing their complexities, BDM estimates the complexity of the whole object. This approach is especially practical when calculating exact algorithmic complexity is too difficult due to the object’s size. By combining the complexity estimates of smaller blocks, BDM offers a broader, more refined view of complexity, surpassing the limitations of basic compressibility measures like Shannon entropy. This method reveals deeper insights into both structured and random data patterns.
    \item z-lib: the length of the b64encode of the compressed zlib version of the string. This is an alternative to LZW lossless compression.
\end{itemize}

Thus, for our initial approach to the problem, the following sequence of steps will be applied:

\begin{itemize}
    \item Consider input string $str_{input}$;
    \item Randomly shift ``\# of flipped bits'' in the string, producing $str_{f,input}$;
    \item Calculate the four metrics previously described for $str_{f,input}$; 
    \item Repeat the previous step 1024 times, keeping the same number of bits flipped but considering different bits each time.
\end{itemize}

In the following subsubsections the results for these perturbations are provided for each type of string. For the violin plots, the ticks inside each violin indicate the quartiles of the data, and the first and last results (no flipping or complete flipping) are represented as circles since the violin plot collapses to a single value distribution.

\subsubsection{Random binary strings}

The first group of strings analysed is the one comprising random binary strings. So, let us consider the following random string of 402 binary numbers:

\begin{eqnarray*}
0001100011110001101101010101010100000100111010100001000110100101000 \\
0101100010010101101011100111010100100000110011101110000001010110010 \\
0001011010001000011001010100010101001101000110111110010010011111010 \\
1110001001111100101100011110000001101010111011011110111100101011011 \\
1001110000011001101100001011100000010000111001101111110110111010111 \\
0100101001111100110010101010001001100100010111011101111111000111011
\end{eqnarray*}

\begin{figure}[!h]
	\centering
	\includegraphics[scale=0.5]{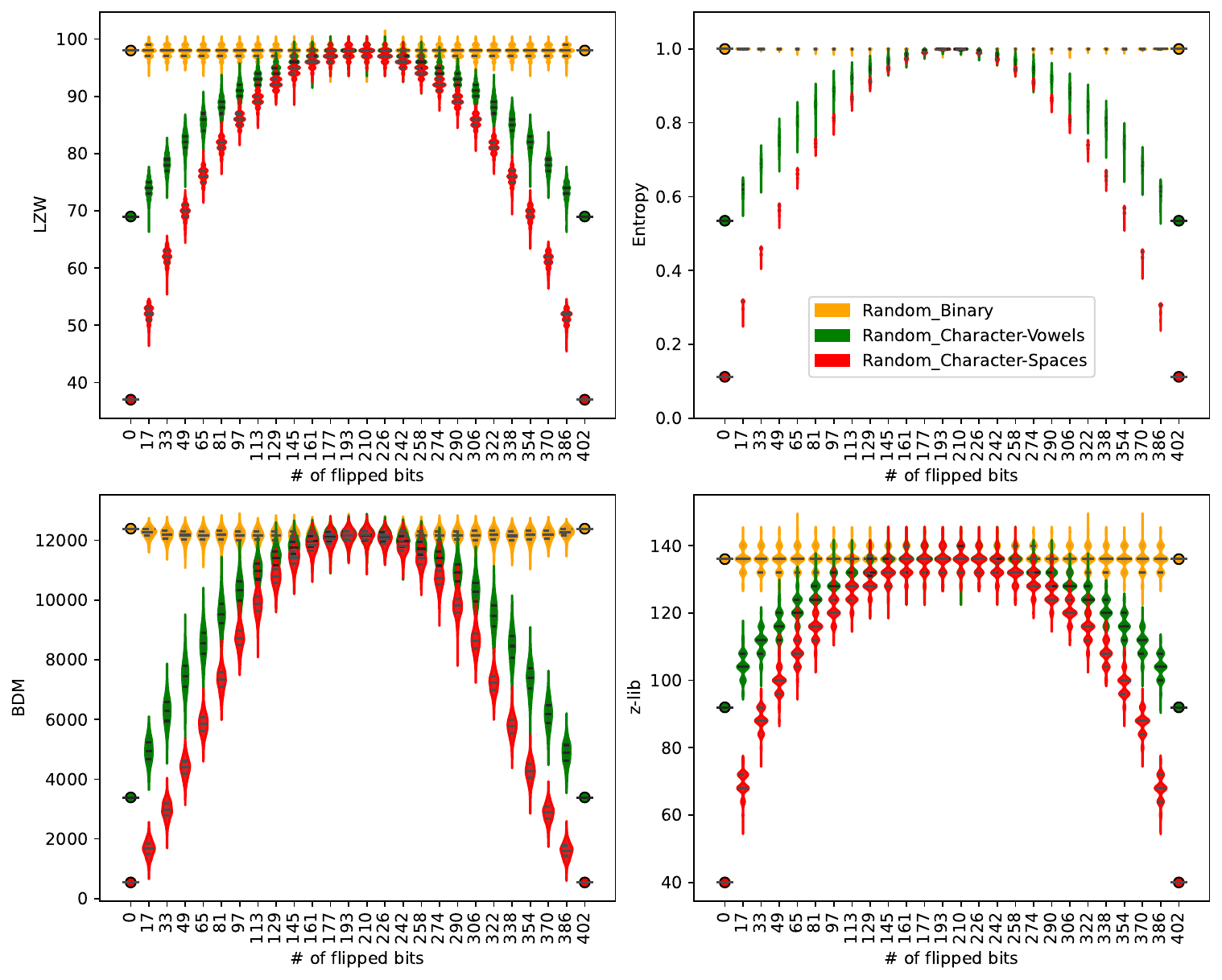}
	\caption{\label{rbin}Algorithmic perturbation analysis on strings of size 402. \textit{Random binary numbers}: For a binary random string with balanced number of 0s and 1s, performing a bit flip for several bits (from 0 - original string, up to 402 - all the characters in the string) causes the resulting size of the LZW dictionary and the zlib encoded string to stay approximately the same (median value constant in the plots). The same is observed for Shannon entropy, which did not change appreciably and stayed close to its maximum value. This is to be expected since our original string is already in a maximal entropy state. On the other hand, BDM slightly decreases, which indicates that the bit flips brought the string to a slightly more informative state (less random too, because, in theory, perfect randomness requires the number of 1s and 0s to be the same, and this is not  guaranteed for the flipped strings). Perturbing balanced binary random messages indicates that, overall, the only information gain possible in the process is obtained when the bit balance is broken, which is graphically represented as an up concavity curve for BDM values. This would be equivalent to tempering a coin and observing a series of its flips. \textit{Random characters sampled from a 95-character alphabet after vowel encoding}: This new binary string is unbalanced, with 12.19 \% of its bits being 1s. Bit flips were performed revealing that all the metrics tend to increase up to when half of the bits are flipped. This is to be expected since our original string, although randomly generated, considered an unbalanced encoding which increased information content. Besides, it is worth noting that the simple fact that the number of 0s and 1s is not the same provides information about the message. Therefore, the down concavity observed implies that the original message is more informative than any perturbation processed. This is because signals carrying meaning are far removed from randomness, and random perturbations make the text more random~\cite{Zenil2019c}, even when the methods know nothing about words, grammar or anything linguistic.\textit{Random characters sampled from a 95-character alphabet after space encoding}: This new binary string is highly unbalanced, with 1.49 \% of its bits being 1s. Bit flips were performed revealing the same behaviour as observed for the vowel encoding. Compared to the vowel encoding, it is possible to see that all the metrics span for a larger interval when highly unbalanced encodings are considered.}
\end{figure}

For such a string, there are exactly 201 0s and 201 1s. By running the simulations indicated, Fig.~\ref{rbin} was obtained. The analysis of Fig.~\ref{rbin} indicates that for binary random strings at maximal entropy, perturbations do not make the string ``more random'', indicating that the information content does not change appreciably after bit flipping. This is a confirmation of the randomness of the input data.

\subsubsection{String of random characters}

For this second class of strings, we will consider strings formed by randomly sampling a string pool. It is possible to consider the first 95 printable characters, for example, and by randomly sampling from this pool, obtain the following string with length 402: \\

{\small
\begin{verbatim}
$eW*yOqM:Lqk#ifo1}o]`404 bx+,QF3JGNh>~#E({z=5W!P3q%$2rn |kZ4eLo33YCN1.O
+iW:_Rjo2rEpd3j?kWw6renR>>E:s,k8t>eG4];YT.t($vWe\'\\e ]TW3YQCZ${ppo(N{h
jBt%m+Pq#tSfr_r\'3FFBy}z7dj)h?(Yi*/pX(i|)\'3@H)@vDvhu<2o&iD)y;F3\'Ivo5f
Zx+:fnYQw8>s) H`bU<MoI/Uh^GGz&`/jy(@Iz)mmH8CV%1U/9ZPOdc.s,">|V7[bd"%(A(
P)-aZ@0=~W(7sR4\'/T~Om~=BCP<y;)1\\d4lufR3A`HG)?4]EZ`Kol|]w*x?*za6Mk6R\'
g/3[E`MBr/a%TU\\l,~/!PZo|KNJBN{~[}a#ZIY)HUZTB+I4? g EQT"

\end{verbatim}
}

\begin{figure}[ht!]
	\centering
	\includegraphics[scale=0.5]{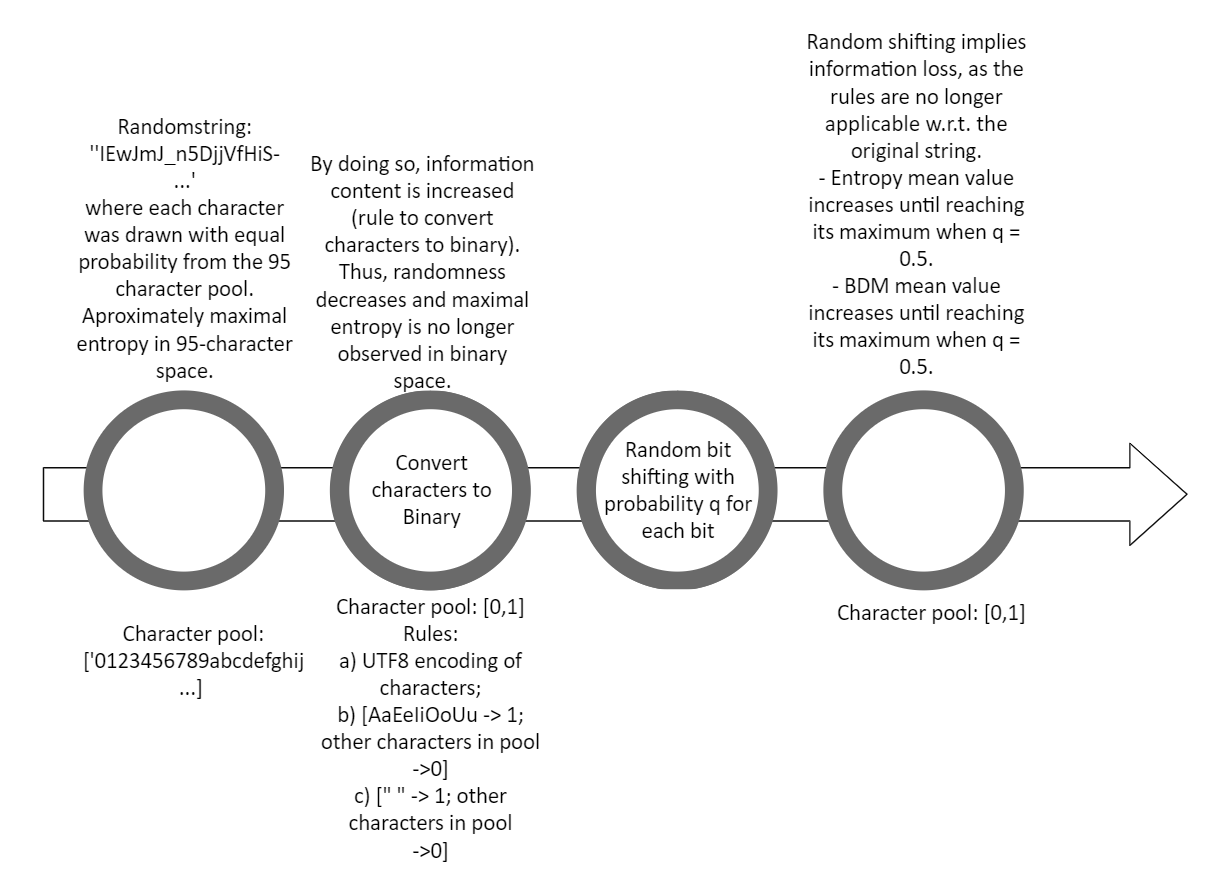}
	\caption{\label{rcharflow}Algorithmic perturbation flowchart for strings of characters randomly picked from a character pool. Even when randomly sampling from a character pool, it is possible that maximal entropy is not observed even in the original alphabet space. By using encoding to convert characters to binary, information content as well as order is introduced in the process, thereby making the converted string ``less'' random than the original one.}
\end{figure}

At the outset it is worth highlighting that even though the string was randomly built, it is not perfectly random since the frequency distribution of characters is not perfectly uniform. This is a direct consequence of the fact that we sampled our original 95-character alphabet 402 times (which means there are more samples than characters, implying oversampling of some of the alphabet elements - impossible to circumvent since 402/95 is not an integer). Therefore, this string does not have maximal entropy over the 95-character alphabet space. Besides, even if we could produce strings with maximal entropy over the 95-character space (for example, a string of size 380 with each character appearing exactly 4 times but in a random order), in general when converted to binary, the new converted string would not have maximal entropy anymore, since the encoding used will increase information content and diminish randomness. Fig.~\ref{rcharflow} presents a flowchart to illustrate this process.

In general, when an encoding algorithm is considered, this implies conditioning the original random sampling, so that a maximal entropy string is almost always ``organised'' when converted to binary. It is interesting to understand how different encodings behave in that context.

\begin{itemize}
    \item Unbalanced encodings
\end{itemize}

By unbalanced we mean encodings which do not preserve the balance of 0s and 1s for each converted character. For example, let us consider the case where each vowel (upper and lower case - AaEeIiOoUu) is converted to 1 while other characters are converted to 0. This encoding is highly unbalanced since, on average, for long random strings taken from the pool, 10/95 of the converted characters will be 1s  while 85/95 will be 0s. Fig.~\ref{rbin} presents the perturbation analysis for the binary string created after applying this encoding. 


It is possible to repeat this same exercise with another, even more unbalanced encoding: each space is converted to 1 while all the other characters are converted to 0. For the string considered, this brings the percentage of 1s to a mere 1.49 \%. The perturbation results are given in Fig.~\ref{rbin}.

\begin{figure}[ht!]
	\centering
	\includegraphics[scale=0.5]{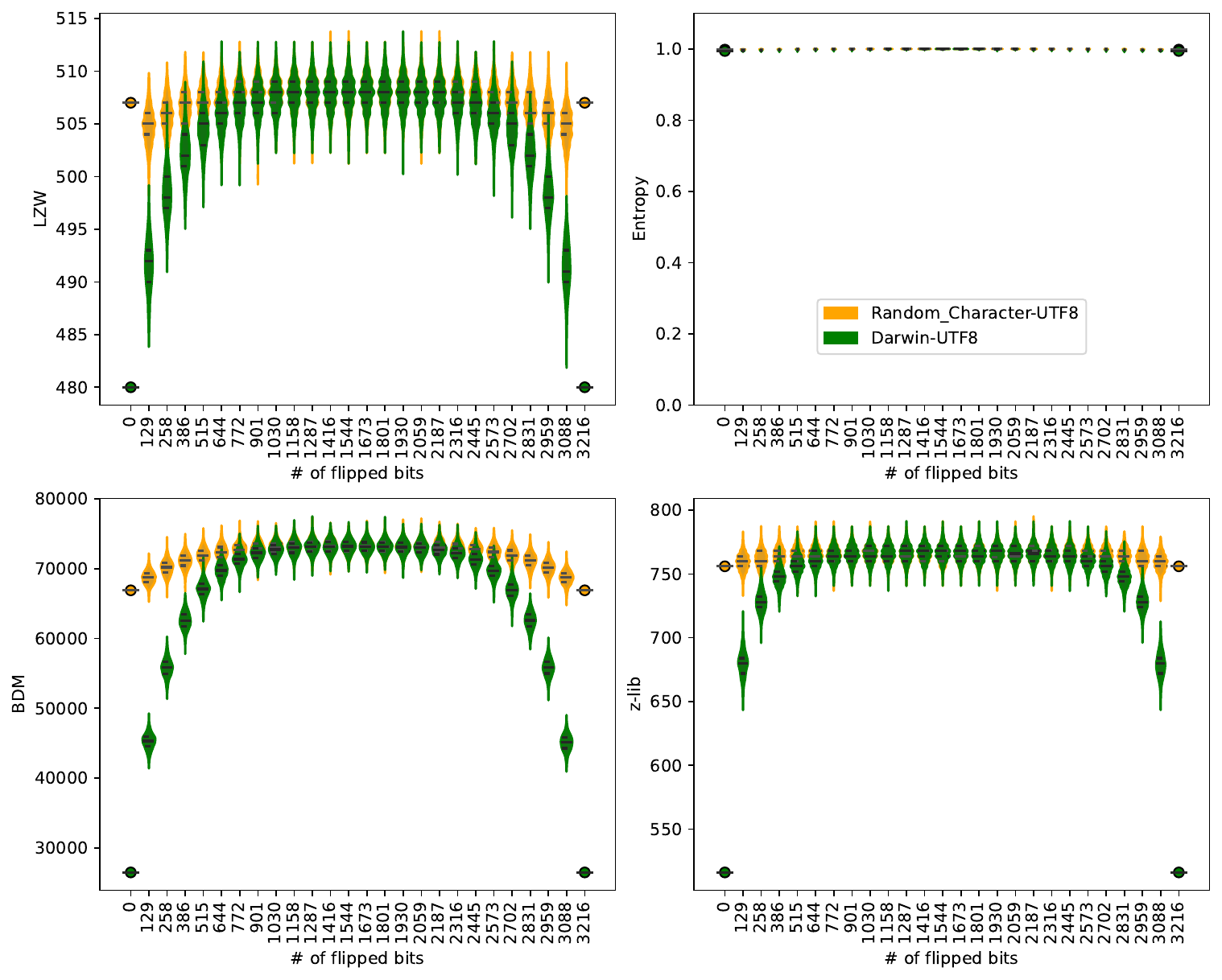}
	\caption{\label{unbrstr3}Algorithmic perturbation analysis on strings of size 3216. \textit{Random characters sampled from a 95-character alphabet after UTF-8 encoding}: This binary string is almost balanced, with 47.42 \% of its bits being 1s. Bit flips were performed for several bits (from 0 - original string, up to 3216 - all the characters in the string), revealing that with the exception of BDM, all the metrics only slightly change (less than 2\%). BDM, on the other hand, points to a continuous increase up to when half of the bits are flipped. Once again, the down concavity observed implies that the original message is more informative than any perturbation processed. Compared to the other unbalanced encodings, as soon as a more balanced encoded string is seen, entropy and compression algorithms tend to lose their power to detect information content changes. \textit{Excerpt of  Darwin's ``Origin of Species'' after UTF-8 (unbalanced) encoding}: This binary string is almost balanced, with 45.55 \% of its bits being 1s. Bit flips were performed revealing that BDM and z-lib were the most sensitive metrics to information content changes. The down concavity observed implies that the original message is more informative than any perturbation processed.}
\end{figure}

If we continue considering unbalanced encodings, it is natural to assess the performance of the traditional UTF-8 encoding. In that case, characters are represented as 8-bit strings, which are then concatenated to form the message to be analysed. Thus, for the initial 402 character message, the binary encoded string is of length 3216. The results for the perturbation analysis of the UTF-8 encoding of the random string of characters is presented in Fig.~\ref{unbrstr3}.

Based on all the examples presented for unbalanced encodings, it is evident that even randomly generated strings may not present maximal entropy when converted to binary. As a matter of fact, as soon as the encoded binary strings start to approach a balance of 0s and 1s, entropy and compression algorithms lose their power to detect information content changes. BDM, on the other hand, presented good results even in more challenging scenarios. Also, if we apply the method to any string and see that the original string (without flipping) is the one with minimum metrics, then the method could not improve on our initial guess about the information content of the message (the message is what it is). This may be true for some 1D strings. On the other hand, for multidimensional cases, there will be downward spikes which will indicate a ``more suitable'' configuration, especially because we may not start the perturbation from the ``correct'' string as we did in the examples above.

\begin{itemize}
    \item Balanced encodings
\end{itemize}

\begin{figure}[ht!]
	\centering
	\includegraphics[scale=0.5]{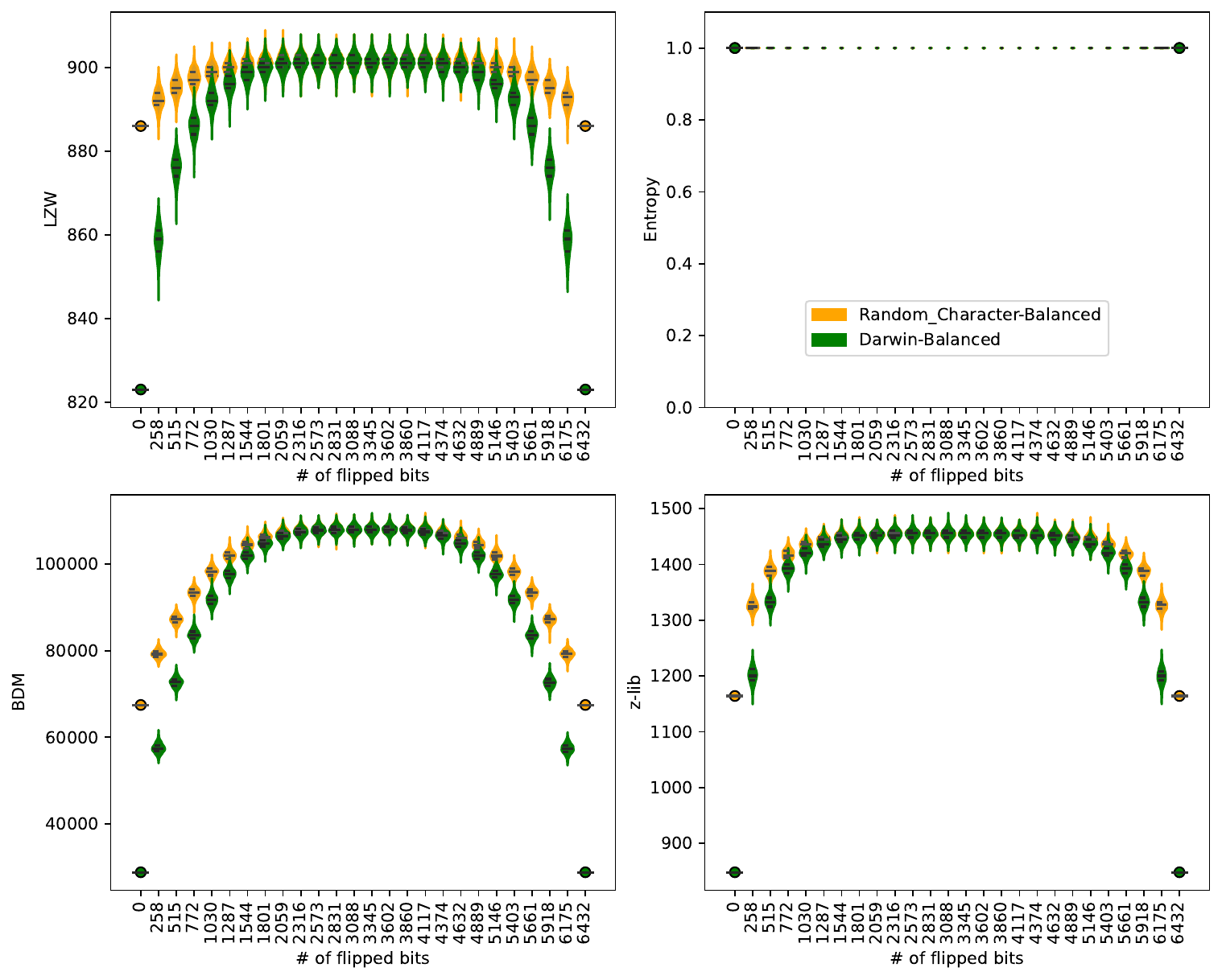}
	\caption{\label{brstr}Algorithmic perturbation analysis on strings of size 6432. \textit{Random characters sampled from a 95-character alphabet after balanced encoding}: This binary string is perfectly balanced, with 50 \% of its bits being 1s. Bit flips were performed for several bits (from 0 - original string, up to 6432 - all the characters in the string), revealing that except from BDM and z-lib, all the metrics present small changes (less than 2\%). BDM and z-lib, on the other hand, present a continuous increase up to when half of the bits are flipped. BDM is more sensitive than z-lib, as the size of the plateau is smaller with longer lateral climbs. Once more, the down concavity observed implies that the original message is more informative than any perturbation processed. Overall, entropy and LZW lose their power as proxies of information content whenever balance is observed for the binary string. \textit{Excerpt of  Darwin's ``Origin of Species'' after balanced encoding}: This binary string is also perfectly balanced, with 50 \% of its bits being 1s. Bit flips were performed, once again revealing the capabilities of BDM and z-lib to assess the information content of the message, and general analysis implies that the original message is more informative than any perturbation processed.}
\end{figure}

In contrast, it is possible to state that a given encoding is balanced if the number of 0s and 1s is balanced (equal) for every character converted. This can be achieved by simply concatenating the UTF-8 encoding of a given character (8-bits) with its bitwise inverted encoding (additional 8-bits), producing a 16bit encoding per character. Such an encoding will always have the same number of 1s and 0s. In this case, the original 402 characters become 6432 bits and the results are presented in Fig.~\ref{brstr}.

\subsection{Sentences from Darwin's ``Origin of Species''}

Following the same rationale, we shall apply the perturbation techniques to encoded binary strings obtained after applying both unbalanced (strict UTF-8) and balanced (UTF-8 concatenated with its bitwise inverted encoding) encodings to the following sentences from Darwin's ``Origin of Species'' (total length of 402 characters):

\begin{lstlisting}[breaklines]
we have many slight differences which may be called individual differences, such as are known frequently to appear in the offspring from the same parents, or which may be presumed to have thus arisen, from being frequently observed in the individuals of the same species inhabiting the same confined locality. No one supposes that all the individuals of the same species are cast in the very same mould 
\end{lstlisting}

The perturbation analyses for unbalanced and balanced encodings of the excerpt from Darwin's ``Origin of Species'' are found in Figs.~\ref{unbrstr3} and \ref{brstr}, respectively.



Figs.~\ref{unbrstr3} and \ref{brstr} reveal that the results observed for random strings are also true for ``real'' sentences: BDM and z-lib are sensitive to information content change and the simpler LZW algorithm together with entropy rapidly lose their abilities to detect information content change whenever the balance of 1s and 0s in the encoded binary string is approached.

The results in this section demonstrate that most single-bit perturbations in a sequence that encodes a message result in a more random-looking sequence. These results vary depending on what mapping we choose to binarise the sequences, but in general, the more compressible the object is, the more a perturbation increases its randomness across various mappings (we also explored other arbitrary binarisation functions, with similar results).
(See also Sup. Mat. and~\cite{Zenil2019c} for mathematical proofs of such a phenomenon). 
In other words, these examples show that sequences that carry meaningful content are quantitatively different according to various measures. Any deviation (i.e., perturbation) from the encoded sequences causes randomness to increase.
Confirming our theoretical results in this paper (see Section~\ref{sectionTheory} and Sup. Mat.), our results suggest that an incoming signal with low randomness is more likely to encode a message than one that has more randomness.

\begin{figure}[ht!]
	\centering
	\includegraphics[scale=0.6]{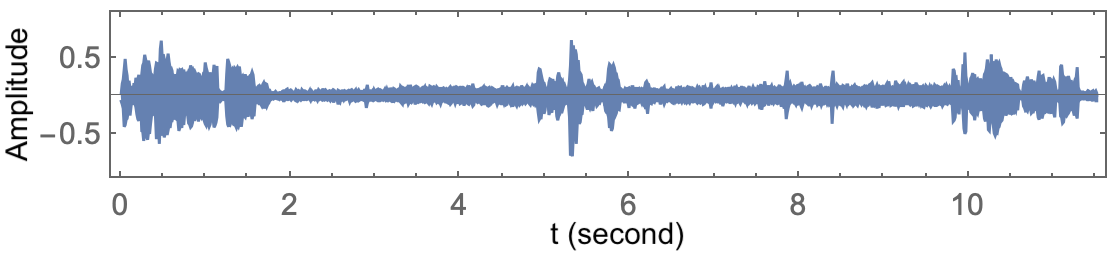}
	\vspace{0.5cm}
	
	\centering
	\includegraphics[scale=0.28]{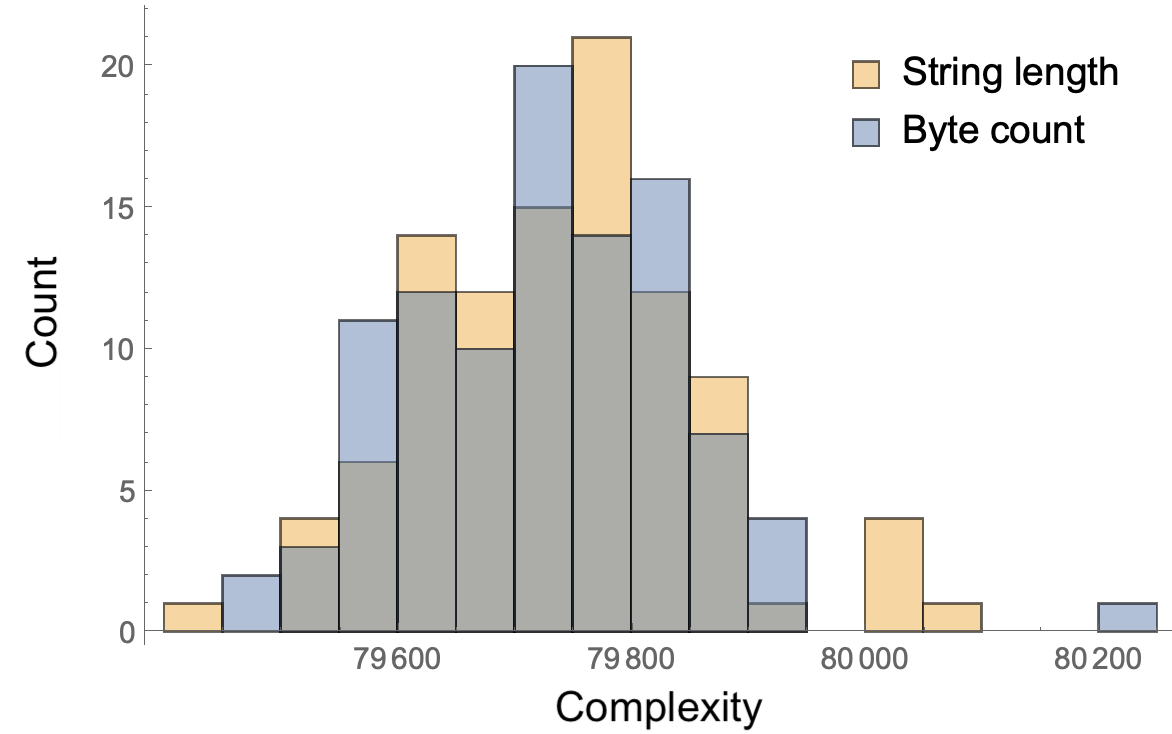}\hspace{0.3cm}\includegraphics[scale=0.31]{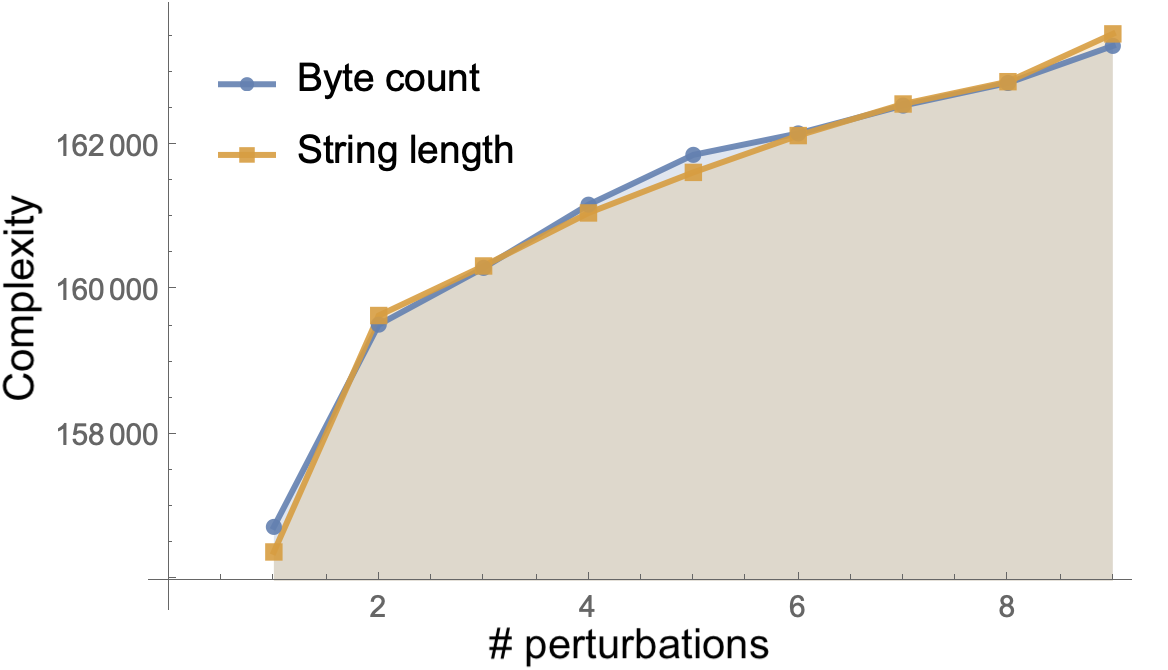}
	\caption{\label{sound}\textit{Top:} Sound waves of the words spoken from the Apollo 13 mission: ``Houston, we've had a problem.''. The message was transmitted on April 13, 1970 at a sample rate frequency of 11025 Hz on a single channel. \textit{Bottom left:} Histograms of the change in compression lengths of scrambled versions of the same message from the length of the original compressed message. By change we mean byte counts minus 106\,250 bytes and string lengths minus 106\,320. \textit{Bottom right:} Small perturbations of the original message by scrambling only the beginning and end of each word shows a smooth transition from low to high randomness. The lowest complexity signal indicates the correct (original) signal. The file was processed in FLAC format (Free Lossless Audio Codec) from a lossless file, with no audio data discarded.}
\end{figure}

Fig.~\ref{sound} demonstrates that the same phenomenon is observed in audio signals, where the perturbation approach has been applied to an audio recording of the words spoken from the Apollo 13 mission:  ``Houston, we've had a problem''. This message was transmitted on April 13, 1970 at a sample rate frequency of 11025 Hz on a single channel. When compressed, the message is 106\,250 bytes and 106\,320 characters long. Here, the message was scrambled several times. The histograms on the bottom left show the difference between compression lengths in bytes and string lengths (length of zlib after b64encoding) of the scrambled messages with respect to the compressed results of the undisturbed message. Each of these scramblings results in an increase in randomness. Additionally, if only the beginning and end of each word is perturbed, the resulting randomness increases smoothly with each perturbation, as shown in the bottom right plot. 
These results indicate that perturbations of the original message, including scrambling elements of the original message in a different order, will result in a message of higher complexity and thus greater randomness. As a result, the original encoding of the message (the one with an interpretable meaning) is the one with the lowest complexity.

Having investigated how bitwise perturbations could point towards better candidates for intelligible messages, it is also interesting to see how rearranging the message could potentially indicate suitable candidates for their respective multidimensional space, such as their natural geometrical features. This was preliminarily explored in~\cite{Zenil2024ETpaper1} and shall be further investigated in the next subsection.

\subsection{Information content and structural perturbations}\label{sectionGeom}

All the examples presented so far were created based on some sort of encoding from a high information content message to binary. Thus, in general, it would be expected that groups of binary elements in the string, not single bits, would carry the actual information content. In this regard, it is interesting to reshape the original binary string into several different configurations, then measure some of the metrics considered during the bitwise perturbations. Considering that z-lib and LZW algorithms are both compression algorithms, we chose to evaluate only z-lib, BDM and entropy.

Since the size of the original binary message of size $s$ may not necessarily be reshaped as $m \times n$ (i.e., the original size is not the product of $m$ and $n$), it is important to set the maximum information loss allowed in this reshaping process. This can be set, for example, as a 1\% loss for the purposes of our investigation, which indicates that we are looking for $m$ and $n$ such that $mn \geq 0.99 s$, where $s$ is the length of the original encoded binary message. Also, there may be more than one pair of $m$ and $n$ values which satisfy such an inequality. 
Thus, we set $n$ as the maximum second dimension for a given $m$ value. 
So, the dimensions actually searched are $(m,n)$, both natural numbers, such that $n=max(n_1,n_2,...n_j)$, where $m n_i \geq 0.99 s$, for $i=1,...,j$.

Besides, we will compare values for which small information loss is possible (since $|s-mn| \leq 0.01$), making it important to take this into account while calculating the metrics. The following changes are then considered:

\begin{itemize}
    \item the BDM value is divided by the number of 4x4 grids used to make the partitioning of the original 2D array (i.e. $(m-3)(n-3)$);
    \item Entropy value is changed to Block Entropy, where the block size is equal to $n$;
    \item zlib b64encoding length is divided by the ratio of bits kept (i.e. $mn/s$).
\end{itemize}

To make the comparisons more clear, for every metric, a MinMaxScaler was applied, such that the results range from 0 to 1.

\begin{figure}[ht!]
	\centering
	\includegraphics[width=0.48\textwidth]{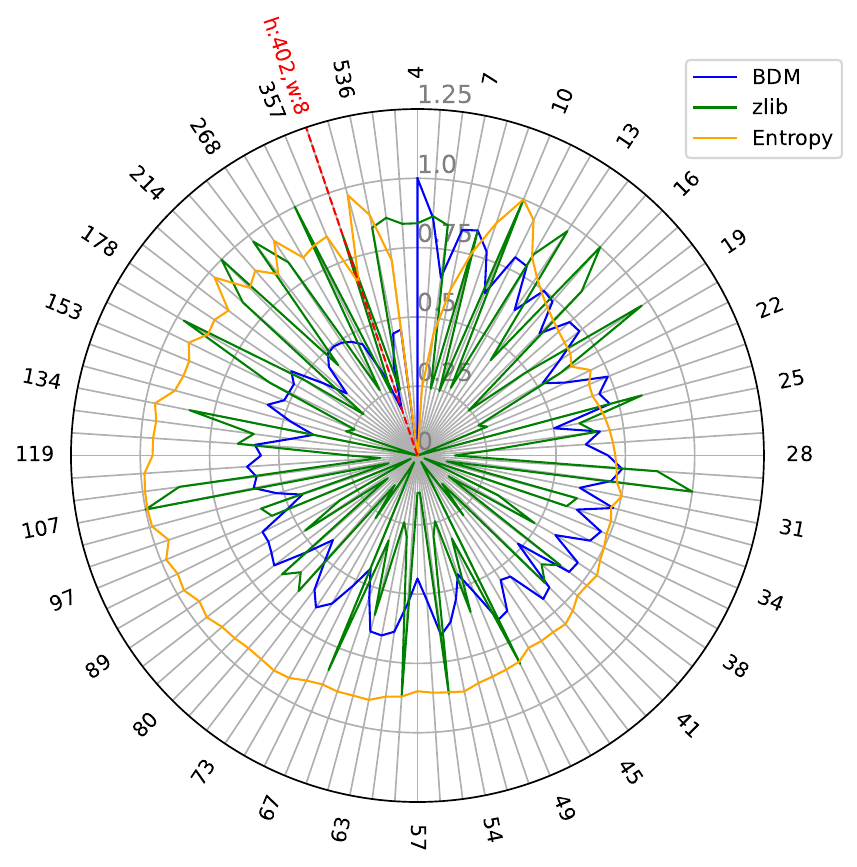}\hspace{0.3cm}\includegraphics[width=0.48\textwidth]{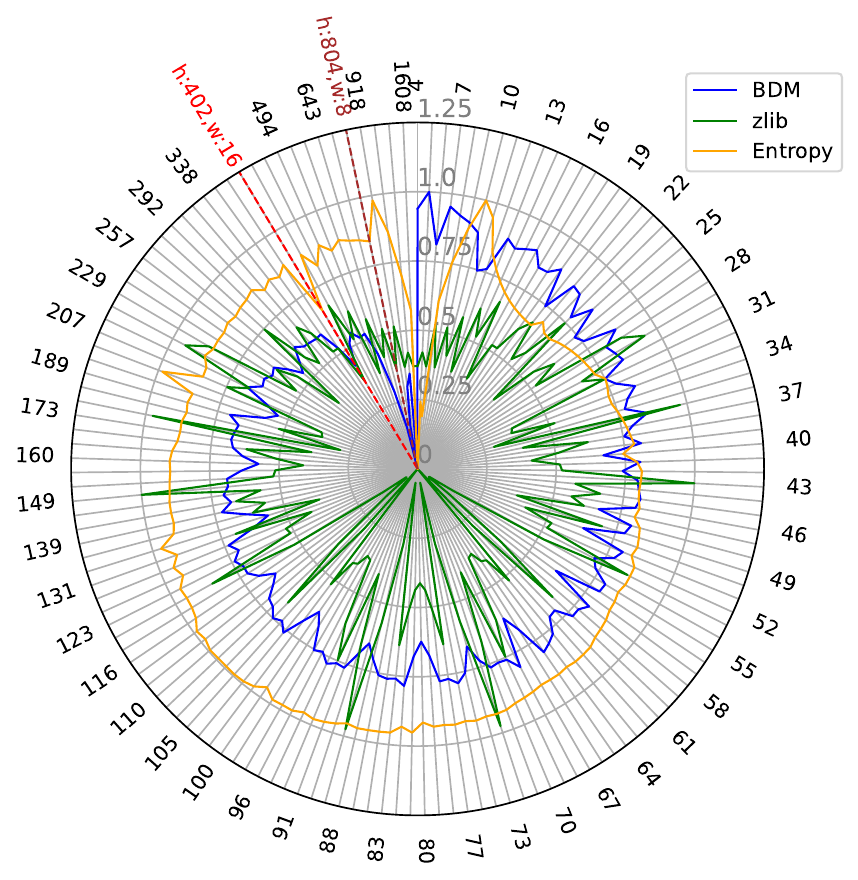}
	\caption{\label{radars}\textit{Left:} Structural perturbations on the UTF-8 binary encoding of Darwin's book excerpt. \textit{Right:} Structural perturbations on the balanced encoding of Darwin's book excerpt. For both cases, the ticks on the radar plot are the values of $m$ obtained after reshaping the initial string as a $m \times n$ array. Also, the ``correct'' shape of the message is represented in red and is observed for downward BDM peaks in the low-BDM region. It can be seen that the balanced encoding boosted the detection of the original UTF-8 encoding (which would be equivalent to a $804 \times 8$ array), since the algorithmic complexity of the bit shifting operation is small.}
\end{figure}

For the excerpts from Darwin's book, the results of this reshaping process are presented in Fig.~\ref{radars}. 
This figure reveals that structural perturbations are able to suggest possible structural arrangements for the message in the space into which it is embedded, allowing one to shortlist a few candidate dimensions based on downward peaks of BDM in the low-BDM region. This is done in a totally agnostic fashion. Since this method showed promising results for text and audio data, it would now be of interest to explore images.

So far, robust empirical evidence has been presented to support the new method hereby proposed. In the next section, theoretical foundations of the method shall be presented.

\section{Methods}\label{sectionMethods}

\subsection{Formalism and basic concepts}\label{sectionBasicconcepts}

For the purposes of the present article, we say an object has ``meaning'' when the object conveyed in the received signal is actually embedded into (or grounded on) the multidimensional space (i.e., the context) that the emitter agent originally intended.
Each (finite and discrete) \emph{multidimensional space} $ \mathcal{ S } $---no matter how complex it is---can be univocally determined by the number of dimensions, the encoding of the set of elements for each dimension, and the ordering in which each dimension appears.
For example: 
the unidimensional space $ \mathcal{ S }_1 $ that takes values from the natural numbers can be univocally determined by only informing the length $ \left| \mathcal{ S }_1 \right| $ in $ \mathbf{O}\left( \log\left( \left| \mathcal{ S }_1 \right| \right) \right) $ bits; 
the bidimensional space $ \mathcal{ S }_2 $ by a pair $ \left( \left| {  \mathcal{ S }_2 }_x \right| ,  \left| { \mathcal{ S }_2 }_y \right| \right) $ in $ \mathbf{O}\left( \log\left( \left| {  \mathcal{ S }_2 }_x \right| \right) \right) + \mathbf{O}\left( \log\left( \left| {  \mathcal{ S }_2 }_y \right| \right) \right) + \mathbf{O}\left( 1 \right) $ bits, 
where $ { \mathcal{ S }_2 }_x $ is the first dimension and $ { \mathcal{ S }_2 }_y $ is the second dimension, and so on.

In this manner,
one is always able to (uniquely) decode the received message into sufficient information to completely determine the multidimensional space configuration.
One example of an encoding scheme for an arbitrary multidimensional space $ \mathcal{ S } $ consists in encoding it in the form of a companion tuple $ \bm{ \tau } $ as in~\cite{Abrahao2020c,Abrahao2021}.
Finite and discrete multidimensional spaces with more intricate configurations other than whole number multiplication in the form of $ m_1 \times m_2 \times \dots \times m_k $, where $ m,k \in \mathbb{N} $, can, for example, be encoded by node-unaligned companion tuples $ \bm{ \tau }_{ ua } $ as in~\cite[Definition~2]{Abrahao2021}.
Nevertheless, the theoretical framework of our method is agnostic vis-a-vis any arbitrarily chosen computational scheme for encoding multidimensional spaces.

Each \emph{partition} of a $n$-bit-length linear signal stream corresponds to a distinct configuration of dimension lengths, and therefore to a distinct multidimensional space in which the total additive dimension lengths remains upper bounded by the constant $ n $.
For example, a picture of 4$\times$4 pixels can be partitioned into other dimension-length configurations leading to different non-squared shapes, like 2$\times$8, 8$\times$2, or 1$\times$16.
Once a row (or, alternatively, a column) is fixed in a bidimensional space, the remaining rows and columns are dependent on the initial row fixation. 
In other words, each bidimensional partition is a single-variable dependency, which is the main rationale underlying the sweep in our algorithms employed to achieve the results discussed in Section~\ref{sectionResults}.

More formally, in the case of a (finite and discrete) bidimensional space $ \mathcal{ S }_2 $, 
one may have different combinations of values $ \left| {  \mathcal{ S }_2 }_x \right| $ and $ \left| {  \mathcal{ S }_2 }_y \right| $ for each partition.
However, for any partition one has it that both
$
1 \leq \left| {  \mathcal{ S }_2 }_x \right| \leq n
$ and
$
\lfloor \frac{ n }{ \left| {  \mathcal{ S }_2 }_x \right| } \rfloor = \left| {  \mathcal{ S }_2 }_y \right|
$
hold.
Thus, because each partition is single-variable dependent on $ { \mathcal{ S }_2 }_x $, for any $n$-bit-length linear signal stream one can encode each of its $ 2 $D-partitions in $ \mathbf{O}\left( \log\left( n \right) \right) $ bits.
The same applies to $ 3 $D-partitions (or to any partition from a finite number of dimensions), except for an extra number of bits upper bounded by a partition-independent constant.
Notice that one can restrict the search for partitions for which the resulting product of $ \left| {  \mathcal{ S }_2 }_x \right| $ and $ \left| {  \mathcal{ S }_2 }_y \right| $ are as close as possible to the original value of $ n $.
For example, this is employed in Section~\ref{sectionGeom} with a constant loss rate, thereby not changing the resulting $ \mathbf{O}\left( \log\left( n \right) \right) $ bits necessary to encode a particular partition in such a restricted search.

As formalised in~\cite{Abrahao2021b} any transformation or perturbation of a finite (encoded) object into another finite (encoded) object is equivalent to an algorithmic transformation that takes the former as input and outputs the latter.
That is, any transformation of a finite object into another finite object is equivalent to an \emph{algorithmic perturbation} $ \mathcal{P} $ (i.e., a program) that takes the former as input and outputs the latter~\cite{Abrahao2021b}.
In this manner, the function $ \mathbf{U}\big( \left< \left< y , \, \mathcal{ S } \right> , \mathcal{P} \right> \big) $ returns the outcome $ \left( y' , \mathcal{ S }' \right) $ of the computational transformation that corresponds to the algorithmic perturbation $ \mathcal{P} $ on the object $ y $ embedded into the multidimensional space $ \mathcal{ S } $ when such transformation is run on the universal machine $ \mathbf{U} $.

A (-n algorithmic) perturbation with low algorithmic complexity is one of these transformations where the algorithmic information content of the original object is mostly preserved under such perturbations.
More formally, a \emph{low-complexity perturbation} occurs when $ \mathbf{K}\left( y \middle\vert x \right) = \mathbf{O}\left( \log\left( \left| x \right| \right) \right) $, where $ \mathbf{K}\left( \cdot \right) $ denotes the algorithmic complexity, $ x $ is the original encoded object, $ y $ is the encoded object that results from the perturbation on $ x $, and $ \left| x \right| $ is the size of the object $ x $.

Swapping \emph{all} 0s for 1s, or swapping \emph{all} black for white pixels in an image are examples of low-complexity perturbations (in these two cases in particular, one has it that $ \mathbf{K}\left( y \middle\vert x \right) = \mathbf{O}\left( 1 \right) $).
Randomly inserting or deleting a finite number of elements (such as creating or destroying edges in a graph or flipping a finite number of bits in a string) are also examples of low-complexity perturbations~\cite{Abrahao2021b,Zenil2019c,algodyn}.
As we saw in the above paragraphs, changing the partition is also a low-complexity perturbation of the multidimensional space into which the object $ x $ is embedded. 
This is because any new partition can always be encoded in $ \mathbf{O}\left( \log\left( \left| x \right| \right) \right) $ bits.
Notice that a perturbation can change:
\begin{itemize}
\item either the object (e.g., by flipping bits), thus called an \emph{object perturbation} or a content perturbation; 

\item or can change its multidimensional space (e.g., by reconfiguring the dimension lengths of a given partition to achieve another distinct partition) independently, thus called a \emph{structural perturbation} or context perturbation; 

\item or can change both at the same time.
\end{itemize}
In any of these three options, except for the identity transformation, the encoded form of the message resulting from the perturbation will be distinct from the encoded form of the original message.
Both changing the partition into which the object is embedded and introducing noise into the object itself (e.g., by flipping bits) are particular examples of (algorithmic) perturbations~\cite{Abrahao2021b,Zenil2019c,algodyn}. 
Notice that these properties of low-complexity perturbations underpin the methodology employed in Section~\ref{sectionPerturbationAnalysis} and Sup. Mat.. 

$ \mathbf{K}\left( \mathcal{ S } \right) $ gives the algorithmic complexity of the arbitrarily chosen encoded form of the multidimensional space $ \mathcal{ S } $.
$ \mathbf{K}\left( y \, \middle\vert \, \mathcal{ S } \right) $ is the algorithmic complexity of the encoded object $ y $ once the multidimensional space $ \mathcal{ S } $ is a priori known.
$ \mathbf{K}\left( y , \mathcal{ S } \right) $ is the algorithmic complexity of the encoding of the object $ y $ univocally embedded into the multidimensional space $ \mathcal{ S } $.
Note that there is no loss of generality in our proofs, nor any algorithmic information distortion (except for an independent constant), in employing here the usual notation $ \mathbf{K}\left( w , z \right) $, which usually stands for the joint algorithmic complexity of a string $ w $ and a string $ z $.
This is because there is an algorithm that can always encode the object $ y $ univocally embedded into the multidimensional space $ \mathcal{ S } $, if the shortest program that generates the encoding of the pair $ \left( y , \mathcal{ S } \right) $ is given as input to this algorithm;
and there is another algorithm that can always return the encoded pair $ \left( y , \mathcal{ S } \right) $, if the shortest program that generates the encoded form of the object $ y $ univocally embedded into the multidimensional space $ \mathcal{ S } $ is given as input to this algorithm.
For example, as shown in~\cite{Abrahao2020c,Abrahao2021}, the composite edge set string of a multidimensional network is always computably retrievable from its characteristic string and the encoded form of its companion tuple;
and both the characteristic string and the companion tuple of a multidimensional network are computably retrievable from the composite edge set string. 

\subsection{Theoretical results}\label{sectionTheory}

Let $ \mathbf{P} \left[ \cdot , \cdot \right] $ be an arbitrarily chosen (computable) probability measure on the space of all possible (computably constructible) finite objects $ y $ embedded into a multidimensional space $ \mathcal{ S } $.
For example, $ \mathbf{P} \left[ y , \mathcal{ S } \right] $ may be any probability measure value that is calculated for $ y $ embedded into $ \mathcal{ S } $ by an arbitrarily chosen formal theory $ \mathbf{F} $, encoding-decoding scheme, and decoding algorithm.

We know from the algorithmic coding theorem~\cite{Chaitin2004,Calude2002,Li1997,Downey2010} that
\begin{equation}\label{equationACT}
	\begin{aligned}
		\mathbf{K}\left( x \right) = \\
	- \log\left( \mathbf{P}_{ \mathbf{U} } \left[ \text{``event } x \text{ occurs''} \right] \right) \pm \mathbf{O}( 1 )
	= \\
	- \log\left( \mathbf{m}\left( x \right) \right) \pm \mathbf{O}( 1 )
	\text{ ,}
	\end{aligned}
\end{equation}
holds,
where:
$ \mathbf{m}\left( \cdot \right) $ is a maximal computably enumerable semimeasure; 
and $ \mathbf{P}_{ \mathbf{U} } \left[ \cdot \right] $ denotes the \emph{universal a priori probability} of an arbitrary event.
$ \mathbf{P}_{ \mathbf{U} } $ can be understood as the probability of randomly generating (by an i.i.d. stochastic process) a prefix-free (or self-delimiting) program that outputs $ x $.
A computably enumerable semimeasure $ \mathbf{m}\left( \cdot \right) $ is said to be \emph{maximal} if, for any other computably enumerable semimeasure $ \mu\left( \cdot \right) $ with domain defined for possible encoded objects, where $ \sum\limits_{ x \in \left\{ 0 , 1 \right\}^* } \mu\left( x \right) \leq 1 $, there is a constant $ C > 0 $ (which does not depend on $ x $) such that, for every encoded object $ x $,
$ \mathbf{m}\left( x \right) \geq C \, \mu\left( x \right)\text{ .} $
Also note that the algorithmic coding theorem applies analogously to the conditional algorithmic complexity $ \mathbf{K}( z \, | w ) $.

We can explore the power and universality of the algorithmic coding theorem in order to investigate a lower bound for the universal (probability) distribution in comparison to an arbitrary probability measure $ \mathbf{P} $. 
A chosen probability measure $ \mathbf{P} $ may, or may not, assign a higher probability value to a correct multidimensional space.
Thus, it is important to evaluate how much a universal approach such as the universal a priori probability can be wrong (i.e., diverges from $ \mathbf{P} $) in the asymptotic limit.
Indeed, the algorithmic coding theorem enables one to guarantee that 
the multiplicative error of 
\[ 
\mathbf{P}_{ \mathbf{U} } \left[ \text{``event }  y \text{ occurs in a given } \mathcal{ S } \, \text{''}\right] 
=
\mathbf{P}_{ \mathbf{U} } \left[ y \middle\vert \mathcal{S} \right] 
\]
with respect to 
\[ \mathbf{P} \left[ y' , \mathcal{ S }' \right] \]
is bounded by the algorithmic probability of the inverse algorithmic perturbation $ \mathcal{P}^{ -1 } $ itself.
This means that the easier, or less complex, it is to retrieve $ y $ embedded into $ \mathcal{S} $ from $ y' $ embedded into $ \mathcal{S}' $ the smaller the multiplicative error of employing $ \mathbf{P}_{ \mathbf{U} } $ instead of the arbitrarily chosen $ \mathbf{P} $.

In the general case (where the multidimensional spaces may be much more complex than the objects themselves), even in the ideal scenario in which we correctly infer the actual object sent by a completely unknown emitter, we are still prone to being wrong about the original dimensions' configurations the emitter intended~\cite{Abrahao2020c,Abrahao2021}.
(See also 
Appendix~\ref{sectionAIDistortions} )
This poses a challenge to straightforward applications of the algorithmic information theory (AIT).
In order to tackle such a challenge, we develop a method for multidimensional reconstruction of received low-dimensional signals in which one does not have access to knowledge about the encoding-decoding scheme or the original multidimensional space.
Theorem~\ref{thm5} demonstrates the existence of sufficient conditions (or assumptions) that allow one to achieve an encoding/decoding-agnostic method for reconstructing the original multidimensional spaces. 
It gives rise to an agnostic method that does not need to assume that one a priori knows that the emitter agent's multidimensional space resembles ours, is as complex as ours, or is isomorphic to the structures we currently deem fundamental in mathematics.

\bigskip
\begin{theorem}[Theorem~$2.10$ in the Sup. Mat.]\label{thm5}
	Let $ y $ be an arbitrary object embedded into an arbitrary multidimensional space $ \mathcal{ S } $  such that the following conditions
	\begin{itemize}
		\item[] \begin{equation}\label{equationStrongcompressibilityofthespace}
			\mathbf{K}\left( \mathcal{ S } \right) = \mathbf{o}\big( \mathbf{K}\left( y \, \middle\vert \, \mathcal{ S } \right) \big)
			\text{ ,}
		\end{equation}
		
		\item[] \begin{equation}\label{equationLogcompressibilityofperturbations}
			\mathbf{K}\left( \mathcal{P} \right) = \mathbf{O}\big( \log\left( \left| y \right| \right) \big)
			\text{ ,}
		\end{equation}
		
		\item[] and
		
		\item[] \begin{equation}\label{equationSupercompressibilityofthemessage}
			\mathbf{K}\left( y \, \middle\vert \, \mathcal{ S } \right) \leq \mathbf{o}\big( \log\left( \left| y \right| \right) \big)
		\end{equation}
	\end{itemize}
	hold, where $ \mathcal{P} $ is any algorithmic perturbation that transforms $ y $ embedded into $ \mathcal{S} $ into another $ y' $ embedded into $ \mathcal{S}' $.
	Let $ \mathbf{P}' \left[ \cdot , \cdot \right] $ be the \emph{uniform} probability measure over the space of all possible outcomes (in the form $ \left( y' , \, \mathcal{ S }' \right) $) of algorithmic perturbations on $ \left( y , \, \mathcal{ S } \right) $ that satisfy Equation~\ref{equationLogcompressibilityofperturbations}.
	Let $ \mathbf{P} \left[ \cdot , \cdot \right] $ be any arbitrarily chosen (computable) probability measure on the space of all possible (computably constructible) objects embedded into a multidimensional space. 
	Then, for any $ \left( y' , \mathcal{ S }' \right) $ distinct from $ \left( y , \mathcal{S} \right) $ satisfying Equation~\ref{equationLogcompressibilityofperturbations}, the following strong asymptotic dominance
	\begin{equation}\label{equationMainresultthm5}
		\begin{aligned}
			\mathbf{P} \left[ y' , \mathcal{ S }' \right]
			=
			\mathbf{o}\big( \mathbf{P}_{ \mathbf{U} } \left[ \text{``event }  y \text{ occurs in a given } \mathcal{ S } \, \text{''}\right] \big)
		\end{aligned}
	\end{equation}
	holds with probability (given by the uniform probability measure $ \mathbf{P}' $) arbitrarily close to $ 1 $ as $ \left| y \right| \to \infty $.
	
\end{theorem}
\bigskip

By applying AID to the problem of communicating messages so that one can perform a perturbation analysis on the received signal, Theorem~\ref{thm5} enables one to exploit the semi-computability of algorithmic complexity while taking advantage of the divergence brought about by the landscape of the approximate complexity values, landscape which results from the perturbations and from which we can find the best candidates to choose.
Individually, the approximation to the complexity value of each object (and multidimensional space) is prone to the unknown error margins that follow from its semi-computability.
However, because each pair object-space results from a perturbation, which is a particular case of algorithm, and each algorithm theoretically has an exact complexity value associated to it, one can retrieve a likely correct answer from a sufficiently large landscape of these perturbations. 

As further discussed in 
Appendices~\ref{sectionZKC} and~\ref{sectionAlgorithmiclikelihood},
we believe the results in this article not only contribute with a general method for interpreting messages in zero-knowledge one-way communication channels, but also establish fundamental connections (and posit open problems) with zero-knowledge proofs and the \emph{algorithmic} counterparts of the maximum likelihood method and the noisy-channel coding theorem.

\subsection{Toward a general theory for zero-knowledge one-way communication}\label{sectionTowardgeneralZKC}

The results presented in this article suggest that AID is capable of harnessing (given the availability of adequate computational resources) some of the algorithmic-informational properties that we demonstrated in ideal and generalised scenarios.
As shown in~\cite{algodyn,algodyn2}, AID allows the investigation of causal analysis~\cite{maininfo,Zenil2019b} and solution of inverse problems~\cite{nmi} by taking advantage of a high convergence rate to the algorithmic probability brought along by BDM, probability values which remain stable across distinct models of computation, most prominently for high algorithmically probable objects.

Our theoretical results show that the method employed to achieve the empirical findings in Section~\ref{sectionPerturbationAnalysis}---see also Sup. Mat.
---is sound with regard to the infinite asymptotic limit when computational resources are unbounded; and it also shows that the method can be generalised to multidimensional spaces in addition to bi- or tridimensional ones.
Our inference method is general enough to remain robust vis-a-vis any arbitrarily chosen alternative method to calculate the probabilities of the events, any arbitrarily chosen encoding-decoding scheme, programming language, or computation model, without the need of assuming some bias in the possible outcomes and prior beliefs.


\emph{Theorem~\ref{thm5}} demonstrates that there are conditions for which decoding the message in the original multidimensional space $ \mathcal{ S } $ chosen from the one in which the message has the lowest algorithmic complexity (i.e., highest algorithmic probability, as assured by the universal distribution) becomes eventually more likely to be correct than decoding the message in a distinct multidimensional space $ \mathcal{ S }' $, with or without noise. 
The underlying idea is that the larger the algorithmic complexity of the closer-to-randomness message candidate in the wrong multidimensional space with respect to the algorithmic complexity of the message in the correct multidimensional space, the larger the (algorithmic) likelihood~\cite{Hernandez2018Algorithmicmutations,Zenil2018bReprogrammabilityChemicalNetworks} of the correct multidimensional space with respect to the likelihood of the wrong closer-to-randomness multidimensional space (see also Appendix~\ref{sectionAlgorithmiclikelihood}).
Thus, finding the multidimensional space that minimises the algorithmic complexity the most, eventually increases the probability of finding the correct one in comparison to the probability of finding the wrong closer-to-randomness one. 
Notice that this holds even if one assumes an arbitrarily different probability measure $ \mathbf{P} $  as starting point.

In this manner, we demonstrate that once sufficient conditions for the compressibility of objects, multidimensional spaces, and perturbations (e.g., noise) are met, one-way communication to the receiver becomes possible even if no prior knowledge about the encoding-decoding scheme chosen by the emitter is known.
One can first simply look into the landscape of all possible partitions and low-complexity noise the receiver has access to while assuming no bias toward particular objects and partitions in this landscape. 
Then, despite having no knowledge about the encoding-decoding schemes and thus a lack of biases, one can infer with probability as high as one wishes that the message with the lowest algorithmic complexity is arbitrarily more likely to be correct than any other arbitrary computable method for assigning probabilities to message candidates.

One of the main conditions for this to hold is that the algorithmic complexity of the object $ y $ (with or without noise) in the wrong multidimensional space should increase (i.e., move the message toward algorithmic randomness) sufficiently fast in comparison to the algorithmic complexity of the message in a correct multidimensional space.
To this end, one can assume a message should contain a large amount of redundancies, and thus should be highly compressible with respect to its linear signal stream length $ \left| y \right| $, such as the case in which $ \mathbf{K}\left( y \middle\vert \mathcal{S} \right) = \mathbf{O}\left( \log\log\left( \left| y \right| \right) \right) $ holds, where $ \mathbf{K}\left( \cdot \right) $ denotes the algorithmic complexity.
In this case, most low-complexity perturbations will transform the object and/or partition by sufficiently moving it toward algorithmic randomness so that the conditions are satisfied.
This condition resembles the ones for the \emph{noisy-channel coding theorem} in classical information theory~\cite{Cover2005}, although our method can be understood as being more general than the classical version because:
it holds for zero-knowledge communication channels, while the classical version requires that the encoding-decoding schemes must be previously known by the emitter and receiver; and
stochastic noise (as in the classical version) is a particular case of randomly generated algorithmic perturbations~\cite{Abrahao2021b}.
See also Appendix~\ref{sectionAlgorithmicnoisychannel}
for further discussion on this and other mathematical phenomena.

The proofs of all theorems can be found in Section~$2$ 
of the Sup. Mat..

\section{Discussion}

We have introduced and demonstrated a general method for reconstructing properties of non-random signals or messages, relating their physical properties to information, entropy, complexity, and semantics.
Our method shows how the receiver can decode the (multidimensional) space into which the original message was sent via zero-knowledge one-way communication channels. 
We have presented signal-amplification techniques that enable the methods to get a better signal-to-noise reading and investigated how the methods perform in the face of noise.

Our method is based on algorithmic information dynamics (AID) which combines tools from classical information theory and algorithmic information theory, in which algorithmic complexity values are approximated and combined with Shannon entropy by using the block decomposition method (BDM), in order to perform a perturbation analysis so that this algorithmic-information-based method can be used to avoid encoding multidimensional distortions or biases in signal detection and processing. 
Our method is asymptotically agnostic vis-a-vis not only \emph{encoding-decoding schemes}, but also to the arbitrarily chosen:
computation model or computational appliances made available;
programming language;  
computable (or semi-computable) method of approximation to the algorithmic information content; 
(computable) probability measure of the events or models;
and formal theory able to formalise the mathematical statements.
Our theoretical framework and results can be extended and used for bio- and technosignature detection and signal deconvolution.
Such results relate information theory to fundamental areas of mathematics, such as geometry and topology, by means of compression and algorithmic probability.

As shown in~\cite{algodyn,algodyn2}, AID allows the investigation of causal analysis~\cite{maininfo,Zenil2019b} and solution of inverse problems~\cite{nmi} by taking advantage of a high convergence rate to the algorithmic probability brought along by BDM.
In addition, although it may seem paradoxical at first glance---because one may initially think that the errors produced by the methods for approximating the algorithmic complexity would be inherited by the perturbation analysis phase itself---our theoretical and empirical results show that the perturbation analysis enables one to overcome the semi-computability limitation in algorithmic information theory, but only when, looking at the whole landscape of complexity values, one estimates with a sufficiently large amount of perturbations effected on the received signal.
This is because the algorithmic complexities of the perturbed signals eventually become sufficiently larger than the complexity of the original message.
Hence, a powerful enough method for approximating complexity, such as BDM, can eventually harness this divergence in order to enable the perturbation analysis to statistically distinguish the correct one (i.e., more compressible) from the wrong ones (i.e., more random).
That is, an approximating method with enough computational resources can produce smaller errors than the difference of actual (although semi-computable) complexity values between the correct candidate and the perturbed ones, even though the exact complexity values remain ultimately unknown to the perturbation analyst (i.e., even though the analyst itself can never be sure it has enough computational resources in the first place).
We argue this is related to zero-knowledge proofs and algorithmic generalised versions of the noisy channel coding and the maximum likelihood principle.
Further discussion of these and other topics can be found in 
Appendices~\ref{sectionAIDistortions}--\ref{sectionAlgorithmiclikelihood}.

\section*{Acknowledgments}

Felipe S. Abrah\~{a}o acknowledges support from the São Paulo Research Foundation (FAPESP), grants $2021$/$14501$-$8$ and $2023$/$05593$-$1$.

{
 }

{
\appendices

\section{Avoiding algorithmic information distortions in arbitrarily complex multidimensional spaces}\label{sectionAIDistortions}

In~\cite{Abrahao2020c,Abrahao2021} it is investigated how much of the irreducible information content of a multidimensional network and its isomorphic monoplex network (i.e., its isomorphic graph) is preserved during such an isomorphism transformation.
In fact, isomorphisms are demonstrated to not preserve algorithmic information in the general case.
The algorithmic information of a multidimensional network might be exponentially distorted with respect to the algorithmic information of its isomorphic (monoplex) network (see~\cite[Theorem~12]{Abrahao2021}).
We refer to an algorithmic information \emph{distortion} when the algorithmic information content of an object (e.g., measured by the prefix algorithmic complexity of an encoded form of the object) in a certain context (e.g., the multidimensional space the object is embedded into) sufficiently differs from the algorithmic information content of the same object in a distinct context.
In this regard, algorithmic complexity \emph{oscillations}~\cite{Li1997} may arise from a series of distortions produced by a large enough collection of distinct contexts into which the same object is embedded.

From the well-known invariance in algorithmic information theory (AIT)~\cite{Chaitin2004,Calude2002,Li1997,Downey2010}, one for example knows that every positive or negative oscillation resulting from changing the computation model (or universal programming language) is pairwise upper and lower bounded by a constant that does not depend on the object output by this pair of machines.
Another well-known example of (plain) algorithmic complexity oscillations are those present in initial segments of infinite strings~\cite{Li1997}.

As shown in~\cite{Zenil2024ETpaper1} (see also Section~$3$ in the Sup. Mat.) 
other examples are the oscillations measured by the plots presented in the figures, 
particularly those image plots for which one has the $x$-axis representing the distinct partitions we estimated the algorithmic complexity of the messages into, respectively.
Notice that the downward spikes one can see in the figures 
are examples of algorithmic complexity (negative) oscillations such that for the decoding problem we are interested in, they indicate the partition that corresponds to the correct multidimensional space the message intended. 

The results in~\cite{Abrahao2020c,Abrahao2021} demonstrate a fundamental limitation that mathematical and computational methods should be aware of in order to properly reconstruct the original multidimensional space from a receiver.
To demonstrate this limitation, take for example received signals that encode (either in $1$D or $2$D signal streams) adjacency matrices of the (monoplex) networks whose isomorphic counterparts are multidimensional.
For any arbitrarily chosen formal theory and decoding algorithm, the theorems in~\cite{Abrahao2020c,Abrahao2021} imply that there is a sufficiently large multidimensional network whose multidimensional space cannot be retrieved from the adjacency matrix of the network by the respective formal theory and/or decoding algorithm.
This occurs because the algorithmic information distortions (due to being exponential) simply outgrow any prediction or output capability of the formal theory together with the decoding algorithm, should the network be given as input~\cite[Theorem 12]{Abrahao2021}.

Such distortions demonstrate that in the general case where the multidimensional spaces may be much more complex than the objects themselves, there is a fundamental limitation for any formal theory with regard to the original multidimensional space to which a received message containing an object embedded into a very simple multidimensional space (e.g., a linear one, such as a signal stream) in fact corresponds~\cite{Abrahao2023bSemanticrobustnessCLPMST}.
This implies that even in the ideal scenario in which we correctly infer the actual object sent by a completely unknown emitter, we are still prone to being wrong about the original dimensions' configurations the emitter intended.
Thus, in addition to the distinctive feature of AID in comparison to classical AIT that we will discuss in Appendix~\ref{sectionAIDinZKOQcommunication}, these algorithmic distortions also pose a challenge to straightforward applications of the algorithmic coding theorem~\cite{Chaitin2004,Calude2002,Li1997,Downey2010},
universal (Solomonoff) induction, or minimum description length methods~\cite{miracle,Li1997}.
If the complexity of the object itself is sufficiently smaller than the complexity of the space (or structure) the object is embedded into, there are cases in which minimising the overall algorithmic complexity (or maximising the algorithmic probability) of the entire message may lead to correctly guessing the object while incorrectly guessing the exact configuration of the (multidimensional) space.
Such a challenge may be even more aggravated in the case of arbitrarily large signal streams which correspond to partial messages made available to the receiver agent so that the respective intended messages will be only completed in larger signal streams~\cite{Abrahao2021darxiv}. 
Intuitively, it is straightforward for us humans to conceive and construct an object, and then to transmit a signal with this object encoded in the message the signal is conveying, so that this object is much more complex than the space-time $4$-dimensional space that we are used to.
However, in the general case either the multidimensional space of a unknown source can be much more complex than humans are used to or the object that the unknown emitter is trying to convey can be relatively simple in comparison to this yet unknown multidimensional space.
Both cases can be affected by the limitations brought about by the aforementioned algorithmic information distortions.

Therefore, in order to circumvent such theoretical limitations, one needs to investigate additional conditions to be met so one can correctly infer the message given that the receiver has no priori knowledge about the source.
In other words, with the purpose of developing methods for multidimensional reconstruction of received low-dimensional signals in which one does not have access to knowledge about the encoding-decoding scheme or the original multidimensional space, one should look for conditions or assumptions that avoid such distortions.
One most obvious approach to avoiding them is to limit the algorithmic complexity of the multidimensional space itself with respect to the algorithmic complexity of the object.
As we have shown in Section~\ref{sectionTheory} (formally demonstrated in Section~$2$ in the Sup. Mat.), a proper formalisation of this intuition is indeed sufficient for achieving an encoding/decoding-agnostic method for reconstructing the original multidimensional spaces. 
Thus, the proven existence of these sufficient conditions in turn highlight the generality and contributions of our present results:
our theoretical framework, method, and results are asymptotically agnostic to a priori biases toward arbitrarily (or subjectively) chosen mathematical foundations such as the multidimensional spaces or the structures we might deem fundamental in the mathematics we (i.e., humans) have developed so far.

\section{Algorithmic information dynamics in communication problems}\label{sectionAIDinZKOQcommunication}

Algorithmic information dynamics (AID)~\cite{algodyn,nmi,Abrahao2021b} consists of a combination of perturbation analysis and both classical information theory and algorithmic information theory in order to solve inverse problems and perform causal analysis by investigating information and randomness dynamics.
It searches for possible generative models, which result from perturbations, and tests whether or not they are compatible with the gathered (or intended) phenomena to be explained (or predicted), while this act of sweeping does not in turn influence the generative models found (i.e., it is a blind search). 
The underlying idea is that a computable model is a causal explanation of a piece of data. 
These models can later be exploited as testable underlying mechanisms and causal first principles~\cite{nmi,maininfo}, regardless of those being stochastic, computable, or mixed processes.

Such an investigation process is based on the universal optimality of algorithmic probability proved in AIT so that algorithmic complexity values are approximated and combined with Shannon entropy by using the block decomposition method (BDM). 
Thus, AID constitutes a formal approach to Artificial General Intelligence that requires a massive production of a universal distribution, the mother of all models~\cite{miracle}, to build a very large (semi-computable) model of computable approximate models.  
The semi-computability of algorithmic complexity (or, equivalently, of the universal distribution) means that one can approximate its exact value from above but never with a known and specified precision.

Our theoretical results in Section~\ref{sectionTheory} show that the method employed to achieve the empirical findings in Section~\ref{sectionResults} and in~\cite{Zenil2024ETpaper1}
, and discussed in Section~\ref{sectionPerturbationAnalysis},
is sound with regard to the infinite asymptotic limit when computational resources are unbounded; and in addition it also shows the method can be generalised to multidimensional spaces in addition to bi- or tridimensional ones.
Thus, our empirical results suggest that AID is capable of harnessing (under feasible computational resources) some of the algorithmic-informational properties that we demonstrated in ideal and generalised scenarios.
As shown in~\cite{algodyn,algodyn2}, AID allows the investigation of causal analysis~\cite{maininfo,Zenil2019b} and solution of inverse problems~\cite{nmi} by taking advantage of 
a high convergence rate to the algorithmic probability brought along by BDM, probability values which remain stable across distinct models of computation, and most prominently for high algorithmically probable objects.

As fully explained in the Sup. Mat. (Section~$1$) and in Section~\ref{sectionMethods}, 
the (algorithmic) perturbations that we investigate in this work can change either the object (e.g., by flipping bits) or its multidimensional space (e.g., by reconfiguring the dimensions' lengths of a given partition to achieve another partition) independently, or can change both at the same time.
The reader is invited to notice that the present work applies AID to the problem of communicating messages by performing a perturbation analysis on the received signal.
Thereafter, the principles of AIT are applied through approximations to algorithmic complexity by using BDM.
These approximations are employed as a performance measure that characterizes the criterion to estimate the correct multidimensional space by distinguishing it from the wrong ones, which are the outcomes of the perturbations.
Therefore, as also raised in Appendix~\ref{sectionAIDistortions}, our present approach, or AID in general, cannot be reducible to a straightforward application of AIT (e.g. by using the algorithmic coding theorem to approximate the optimal distribution or another method to approximate algorithmic complexity~\cite{zenilreview2020}) so as to estimate the irreducible information content of the received signal in order to directly retrieve the original message sent by the emitter agent.
This is because it is in fact the semi-computability that evinces the divergence power brought about by the \emph{landscape} of the approximate complexity values, landscape which results from the perturbations and from which we can find the best candidates to choose.
One could empirically estimate the values of the algorithmic complexities of each possible message and also compare them with each other in order to check which one displays the lowest algorithmic complexity, therefore more likely to be the correct one.
However, due to the uncomputability of algorithmic complexity (or algorithmic probability), this approach alone will never assure that the estimation error is not necessarily high enough to make the complexity value of a message actually be lower than the complexity value found in another message, whose approximate complexity value initially displayed a lower value in comparison to that of the first message in the empirical estimation process.
In other words, the uncomputable errors in any estimation of algorithmic complexity can deceive one into thinking a message which actually has higher complexity than another message, has a lower complexity than the latter.
The objects in other dimensions may not actually be more random, they only appear more random to the approximating methods, but it is the perturbation analysis that helps us to find the intended patterns.

Although it may seem paradoxical at first glance---because one may initially think that all the errors produced by the approximating methods to the algorithmic complexity would be inherited by the perturbation analysis phase itself---, it is the perturbation analysis that enables one to overcome such a semi-computability limitation but only when looking at the whole landscape of complexity values one estimates with a sufficiently large amount of perturbations effected on the received signal.
Enabled by the employment of AID in the problem of interpreting a signal without prior knowledge of its source, we believe this also is one of the contributions of our results.
As the number of perturbed messages strongly dominate the number of possible best message candidates, our theoretical and empirical results show that perturbation analysis enables the downward spikes displayed in the landscape of complexities to more likely indicate the best message candidates, even though the exact values of the algorithmic complexities are uncomputable and one can only asymptotically approximate them from above.

The underlying reason this occurs is because the algorithmic complexities of the messages resulting from the algorithmic perturbations eventually become sufficiently larger than the algorithmic complexity of the original message.
That is, the set of more random candidates becomes much larger than the set of compressible candidates.
Hence, a powerful enough approximating method to the algorithmic complexity, such as BDM, can harness this divergence in order to enable the perturbation analysis to statistically set apart the correct one (i.e., more compressible) from the wrong ones (i.e., more random).
This holds because an approximating method with enough computational resources can produce smaller errors than the actual difference of complexity values between the correct candidate and the perturbed ones, even though the exact complexity values remain ultimately unknown to the perturbation analyst (i.e., even though the analyst itself can never be sure it has enough computational resources in the first place).
Individually, the approximation to the complexity value of each object and multidimensional space is prone to the unknown error margins that follow from the semi-computability.
However, because each pair object-space results from a perturbation which is a particular case of algorithm and each algorithm theoretically has an exact complexity value associated to it, one can retrieve a likely correct answer from a sufficiently large landscape of these perturbations. 

Thus, although there is an intrinsically subjective ``deficiency'' in the perturbation analysis that forbids an arbitrarily precise approximation to algorithmic complexity, this subjective aspect can be eventually cancelled out in the perturbation analysis phase.
In case one chooses another complexity measure that is in principle computable instead of the measures given by algorithmic information theory, to be employed in the analysis phase, then Equation~\ref{equationMainresultthm5} would \emph{not} hold in general, and in principle~\cite{zkpaper,zenilreview2020}.
Therefore, the subjective aspects, such as the arbitrarily chosen programming language, probability measure $ \mathbf{P} $, or the computational appliances made available, would not be able to be overcome in the perturbation analysis phase.
In this sense, our results show that it is the very subjective ``deficiency'' given by the semi-computability that counter-intuitively enables the perturbation analysis to eventually overcome this and the other subjective aspects, that is to find an objective piece of information.


\section{Toward an algorithmic noisy-channel coding theorem}\label{sectionAlgorithmicnoisychannel}

The reader is invited to notice that Theorem~\ref{thm5} demonstrates an algorithmic-informational version of the \emph{noisy-channel coding theorem} in classical information theory~\cite{Cover2005}.
Instead of the purely stochastic noise to which the communication channel may be subject in the classical version, the communication problem we studied in this paper deals with communication channels under \emph{algorithmic noise}, i.e., the case in which signals are prone (or may be subject) to \emph{algorithmic perturbations}~\cite{Abrahao2021b} that can change the originally sent signals like a computer program could do, then resulting in new signals---which may substantially differ from the respective original signals.   
This new version of the problem of communicating a message can be understood as being more general than the classical version because:
it holds for zero-knowledge one-way\footnote{ Notice that the classical version also applies to one-way communication but since both agents must know each other's encoding-decoding scheme beforehand, one may argue that at least some information was previously communicated between the emitter agent and the receiver agent, which by definition violates the condition of one-way communication in the first place.}
communication channels, while the classical version requires that the encoding-decoding schemes must be previously known by the emitter and receiver; and
stochastic noise (as in the classical version) is a particular case of randomly generated algorithmic perturbations~\cite{Abrahao2021b}.

However, in addition to Theorem~\ref{thm5} requiring unbounded computation time so that one can calculate the optimal values of algorithmic complexity, one in fact only proves in Theorem~\ref{thm5} sufficient conditions for enabling the (one-way) communication in channels under algorithmic perturbations (i.e., under algorithmic noise, which subsumes but differs from the stochastic noise in the classical version) and zero knowledge about the encoding scheme chosen by the emitter. 
Instead, for a complete analogous to the classical version, one still needs to demonstrate necessary conditions, that is, to investigate tighter bounds for the asymptotic dominances regarding the relationship between the algorithmic complexity of the algorithmic perturbations and the algorithmic complexity of the message itself.
Thus, while we have demonstrated sufficient conditions so that an \emph{algorithmic noisy-channel coding theorem} can hold, future research is paramount for investigating its necessary conditions, like achieving an algorithmic analogous to the \emph{channel capacity}, which in the classical version plays the role of a threshold for enabling or \emph{not} the communication (with errors as low as one wishes).

\section{Zero-knowledge communication and ZKPs}\label{sectionZKC}

As discussed in Appendix~\ref{sectionAlgorithmicnoisychannel}, one of the salient advantages of our results in Theorem~\ref{thm5} is the absence of a condition stating that the encoding-decoding scheme arbitrarily chosen by the emitter agent must be also known by the receiver agent.
Such a condition that relaxes the need of prior knowledge is what defines the \emph{zero-knowledge} characteristic of the problems we investigated in this work.
Our results are not only agnostic to prior knowledge of encoding-decoding schemes used to encode the message by the emitter agent, but also to:
the computation model and
the programming language chosen in order to implement our proposed method;  
the computable (or semi-computable) method of approximation to algorithmic complexity; 
any arbitrarily chosen (computable) probability measure of the events;
and the formal theory able to formalize the mathematical statements.
We have demonstrated sufficient conditions in this zero-knowledge scenario for enabling the reconstruction of the original \emph{message} (i.e., the original object embedded into the original multidimensional space to be conveyed to the receiver) in one-way communication channels.

In this manner, \emph{zero-knowledge communication} (ZKC) occurs when the receiver agent is able to correctly interpret the received signal as the originally intended message sent by the emitter agent given that the receiver has no knowledge about the encoding-decoding scheme chosen by the emitter.
Notice that this condition of no prior information about the emitter agent only applies to encoding-decoding schemes (and programming languages and computation models), and consequentially also to the original multidimensional space and object that are yet unknown by the receiver agent before any communication takes place.
However, this zero-knowledge condition does not mean other assumptions regarding the unknown emitter are not being considered by the receiver (in addition to those mathematical conditions already investigated in Section~\ref{sectionTheory}).
For example, as it is the case studied in the present work, the assumption that the emitter agent is somehow capable of performing an arbitrary encoding and compression of the message into a signal stream.
A necessary assumption is that the signal sent to the receiver actually contains a message the emitter agent is trying to convey.
To also ensure the conditions investigated enable our theoretical results to hold is that the receiver agent also needs to assume the emitter agent is not in fact trying to deceive the other end of the communication channel~\cite{Abrahao2021darxiv}, as mentioned in Appendix~\ref{sectionAIDistortions}.

Not only because of the nomenclature, but also because of a mathematical proposition being ascertained with probability arbitrarily close to $1$ as a larger number of events happen, one may be tempted to relate ZKC with \emph{zero-knowledge proof} (ZKP) as traditionally known in cryptography~\cite{Buchanan2022Cryptographybook,Vadhan2023surveyonZKproofs,Allender2023AITstatisticalZK}.
Indeed, one may consider that they share some features or intuition in common.
For example, notice that the \emph{one-way} characteristic in ZKC problems plays the role of a direct analogous to non-interactive ZKPs.
Nevertheless, it is important to also notice that ZKP and ZKC are based upon \emph{opposing} assumptions and goals regarding the ``zero-knowledge'' characteristic.

In the ZKP case, one looks for high enough probabilities of one agent correctly attesting the other agent's achievement while no information (or knowledge) is revealed about the actual content of this achievement (e.g., the proof of a theorem).
Intuitively, one agent (i.e., the prover) should statistically convince the other agent (i.e., the verifier) of having certain information (i.e., knowledge) in the long run, and this convincing should happen in such a way that the second agent acquire no information about what the former managed to convince the latter.

In the ZKC case instead, one looks for high enough probabilities of one agent being correct about the very \emph{content} of what the other agent intended to communicate, given that no information was known by the former about the latter.
Intuitively, one agent (i.e., the receiver agent) should statistically (algorithmically or universally) guarantee that its own interpretation of the signal sent by the other agent (i.e., the emitter agent) to the former corresponds to the original message that the latter intended, and this understanding (i.e., decoding) should happen in such a way that no prior information (i.e., knowledge) about what and how the second agent will communicate in the first place is known.

On the one hand, the resulting process in ZKPs should lead to no knowledge being acquired as a goal, once the acquisition of any knowledge is undesirable.
On the other hand, the resulting process in ZKC should lead to maximum knowledge being acquired as a goal, once zero knowledge was a starting condition in the first place.
In this sense, one may consider ZKP and ZKC as akin mathematical problems, but diametrically opposing counterparts of each other regarding the acquisition of knowledge.

\section{A perturbation analysis-based method for the maximum algorithmic likelihood}\label{sectionAlgorithmiclikelihood}

In order to tackle the new challenges presented in Appendix~\ref{sectionAIDistortions}, we have investigated additional conditions that enable one to infer the correct message, given that the receiver has no prior knowledge about the source (i.e., object and multidimensional space).
In this sense, one aspect of the contributions of the present work resides in demonstrating the existence of sufficient conditions that allows the application of the \emph{algorithmic coding theorem}~\cite{Calude2002,Downey2010} to estimate the best model or parameters, even in scenarios where algorithmic information distortions may take place due to isomorphic transformations of the multidimensional space.
Being one of the central results and powerful tools from algorithmic information theory, the usage of the \emph{universal distribution} established by the algorithmic coding theorem underpins methods based on universal (Solomonoff) induction or minimum description length~\cite{miracle,Li1997}, whose general solutions encompass and refine the \emph{maximum likelihood} or \emph{maximum entropy} methods~\cite{Cover2005,Li1997,Zenil2019b}.

As discussed in Appendix~\ref{sectionAIDinZKOQcommunication}, another aspect of our empirical and theoretical results resides in showing how perturbation analysis can be employed to overcome some of the semi-computability limitations in the approximations to algoritmic complexity (or algorithmic probability) values.
Because the algorithmic complexities of the perturbations dominate those of the best candidates, one can still statistically infer the most likely messages (associated with higher algorithmic probabilities) due to the increasing divergence between the actual complexity values of these best candidates and the perturbed signals.
By taking the algorithmic complexity of the object and the multidimensional space as the parameters to be estimated, our methods and results demonstrate how one can statistically infer the parameters with highest \emph{algorithmic likelihood}~\cite{Hernandez2018Algorithmicmutations,Zenil2018bReprogrammabilityChemicalNetworks} from perturbation analysis without the need of exactly calculating the (uncomputable) algorithmic complexity values.
In this regard, as demonstrated in Theorem~\ref{thm5}, this inference method is general enough to remain robust vis-a-vis any arbitrarily chosen alternative method to calculate the probabilities of the events, while assuming no bias in the priors (i.e, while taking a uniform distribution of the possible messages as starting point).

For example, Theorem~\ref{thm5} shows that for any probability distribution, encoding-decoding scheme, programming language, and computation model, the algorithmic complexity of the correct message is strongly dominated by the algorithmic complexities of the perturbations, dominance which holds with (uniform) probability arbitrarily close to $1$.
If one knows how to calculate an upper-bound approximation to the algorithmic complexities (approximations which can be calculated by computable methods), then there are sufficient conditions for which one can in general infer that the algorithmic complexity of the sent message increases much slower than the algorithmic complexity of the perturbations, or equivalently that the algorithmic probability of the sent message increases much faster than algorithmic probability of the perturbed signals. 

Intuitively, as it is already known to hold analogously for \emph{Fisher information}~\cite{Cover2005} (which gives a lower bound for the variance of any unbiased estimator) in traditional statistics, sharper and more isolated valleys (i.e., downward spikes) found in the landscape of complexity values calculated for each parameter should indicate a better approximation to the actual algorithmic complexity (or algorithmic probability) of the correct parameter.
Formally, if the algorithmic complexities of the perturbed signals are known to be sufficiently higher than the approximations to the complexities of the best message candidates (along with the fact that the quantity of distinct perturbations also strongly dominates the quantity of best candidates), then Theorem~\ref{thm5} shows that the errors in these approximations are strongly dominated by (i.e., increase much slower than) the highest algorithmic complexity of the perturbations.
Thus, finding the existence of more tighter bounds in the relationship between the variance of the algorithmic complexity change rate (e.g., a rate obtained by calculating the ratio between the information difference~\cite{Zenil2019c} and the algorithmic complexity of the perturbation that changed the elements of the system) and how small those errors are is a problem that demands future investigation.
In contrast to traditional universal induction methods in AIT that looks for the optimal solution (or model) by directly maximising the algorithmic probability of the models (or minimising the complexity of the probability distribution that best fits the data), we believe that our present work also contributes by establishing first results that posit a new path of future research in this direction of investigating the global shape of the landscape of complexities/probabilities of alternative events or models.

}

\bibliographystyle{IEEEtran}
\bibliography{biblio-ETpaper2-IEEETIT-2024}

%
%


\vfill

\includepdf[pages=-]{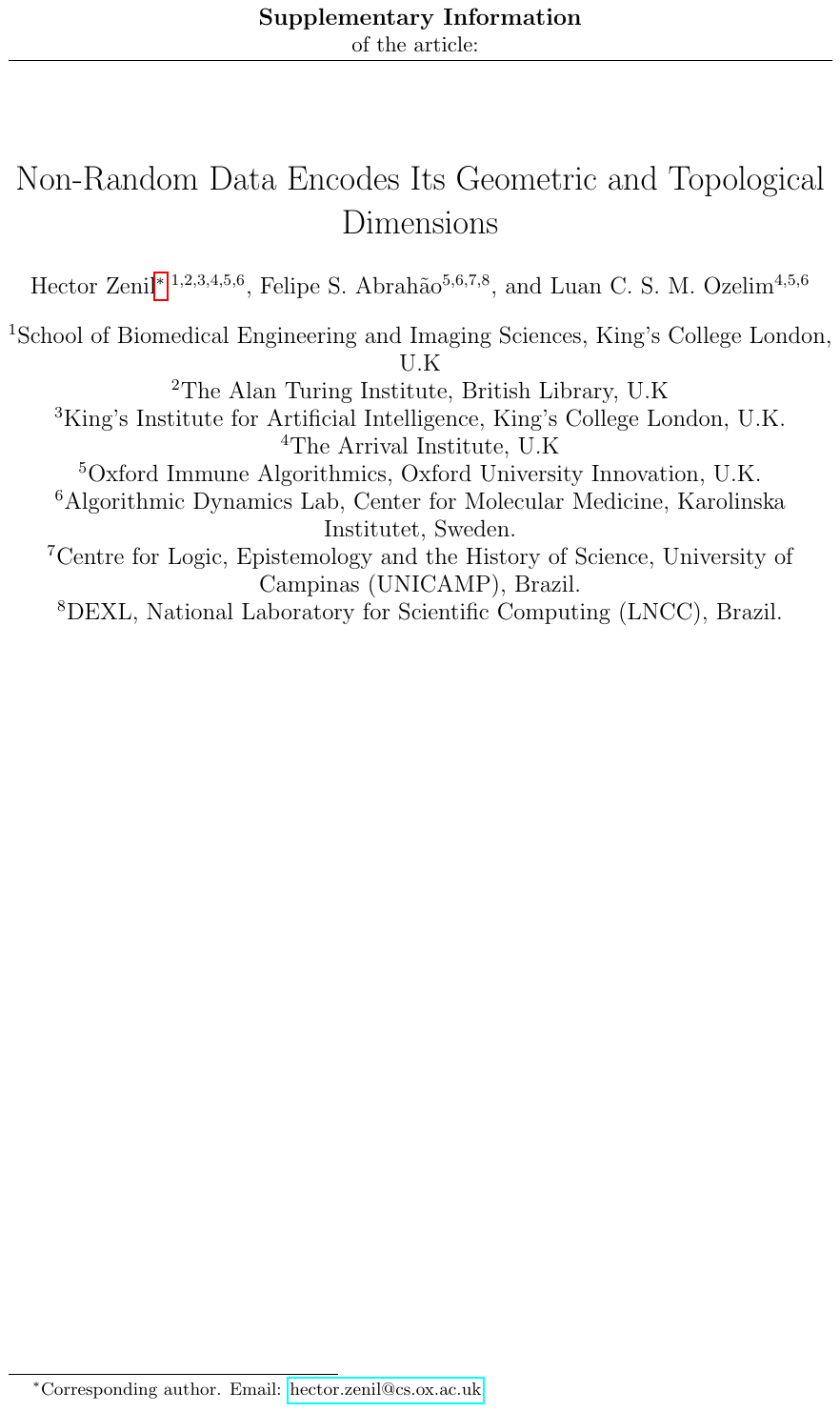}

\end{document}